\documentclass[12pt]{iopart}
\usepackage{graphicx}
\begin{document}

\title{Trends in Heavy Fermion Matter}

\author{J. Flouquet$^1$, D. Aoki$^1$, F. Bourdarot$^1$, F. Hardy$^2$, E. Hassinger$^1$, G. Knebel$^1$, T. D. Matsuda$^{1,3}$, C. Meingast$^2$, C. Paulsen$^4$, V. Taufour$^1$}
\address{$^1$ CEA-Grenoble, INAC/SPSMS, 17 rue des Martyrs, 38054 Grenoble cedex 9, France.}
\address{$^2$ KIT, Institut f\"ur Festk\"orperphysik, 76021 Karlsruhe, Germany.}
\address{$^3$ Advanced Science Research Center, JAEA Tokai 319-1195, Japan.}
\address{$^4$ Institut N\'eel, CNRS/UJF Grenoble, 38042 Grenoble cedex 9, France.}
\ead{jacques.flouquet@cea.fr}
\begin{abstract}

A brief review on major advances in heavy fermion physics is presented including the Ce metal phase diagram, the huge effective mass detected in CeAl$_3$, and the successive discoveries of unconventional superconductivity in CeCu$_2$Si$_2$  and three U based compounds, UBe$_{13}$, UPt$_3$ and URu$_2$Si$_2$. 
In order to track the origin of the huge effective mass, 
the case of intermediate valence compounds is discussed with emphasis of the differences between Yb and Ce materials. The formation of the effective mass is analyzed by two regular- and singular-part contributions.
Examples are given for both, antiferromagnetic (CeRu$_2$Si$_2$ series) and ferromagnetic tricriticalities (UGe$_2$). Pressure and magnetic-field studies on the ferromagnetic superconductor URhGe illustrate the role of the singular effective mass enhancement on the superconducting pairing. 
The discovery of the Ce-115 material gives the opportunity to study deeply the interplay of antiferromagnetism and superconductivity. This is clearly demonstrated by field re-entrance AF inside the SC phase  just below the superconducting upper critical field ($H_{\rm c2}$) for CeCoIn$_5$ or on both side of $H_{\rm c2}$ within a restricted pressure window for CeRhIn$_5$. 
The present status of the search for the hidden-order parameter of URu$_2$Si$_2$ is given and we emphasize that it may correspond to a lattice unit-cell doubling which leads to a drastic change in the band structure and spin dynamic, with the possibility of competition between multipolar ordering and antiferromagnetism.
\end{abstract}

\maketitle

\section{Introduction}
Heavy fermion compounds (HFC) are of special interest as a large variety of ground states can be achieved and a strong interplay between them occurs. 
Due to the weakness of their corresponding characteristic temperature a change from one state to another can be easily realized by moderated temperature ($T$), pressure ($P$), or magnetic field ($H$) tunings. Heavy-fermion quasiparticles are the result of strong magnetic and valence fluctuations.

Heavy fermion physics started with the discovery of the Ce metal phase diagram in which the occupation number ($n_f$) of the 4f trivalent Ce$^{3+}$ configuration is a key parameter~\cite{Jay65}. 
A boost in the field was the discovery in CeAl$_3$~\cite{And75} that even with $n_f$ close to unity 
the establishment of long range magnetic ordering can be avoided with the benefit of the formation of huge heavy quasiparticles.
The Sommerfeld coefficient $\gamma \sim 1\,{\rm J\,mole^{-1}K^{-2}}$ surpasses that of a free electron value by three order of magnitude. 
The description from single Kondo impurity to a regular array of Kondo centers (the Kondo lattice) is still the subject of debate~\cite{Var06}. 
At least, now it is well established that heavy quasiparticles move along the trajectories of the Fermi Surface~\cite{Tai88,Set07}.

A major breakthrough was the discovery of superconductivity (SC) in CeCu$_2$Si$_2$~\cite{Ste79} which opened the era of unconventional SC. 
This new research field was reinforced by the concomitant discoveries of SC in three uranium HFC namely
UBe$_{13}$~\cite{Ott83}, UPt$_3$~\cite{Ste84_UPt3}, and URu$_2$Si$_2$~\cite{Sch86,Pal85}. 
Unconventional SC with a sign-reversing order parameter (OP) was directly proved by the observation of multiple SC phases in UPt$_3$~\cite{Fis89_UPt3}. 
New features of unconventional SC, e.g. the unitary treatment of impurities~\cite{Cof85,Pet86}, were rapidly derived from HFC studies~\cite{Rav87,Jac87},
whereas it took, for instance, a decade before serious considerations of unconventional SC were put forward in organic conductors.

The achievement of huge effective mass was rapidly related to the proximity to a magnetic instability, 
but it took over a decade before the advent of quantitative studies focusing on quantum criticality such as the switch from paramagnetic (PM) to antiferromagnetic (AF) ground states. 
Since this point is highly discussed by other authors~\cite{Col06,Loh07}, 
we will focus here on different facets of HFC going from the main discoveries to the present status.

In this article special attention is given to: (i) the ($H, P, T$) phase diagram ranging from   intermediate valence compounds to examples of tricriticality for both antiferromagnetic (AF) and ferromagnetic (FM) materials, (ii) recent investigations on the interplay between effective mass enhancement and SC, and (iii) the resolution of the puzzle of the hidden order (HO) phase of URu$_2$Si$_2$.


\section{Intermediate valence compounds}

\begin{figure}[tbh]
\begin{center}
\includegraphics[width=0.5 \linewidth]{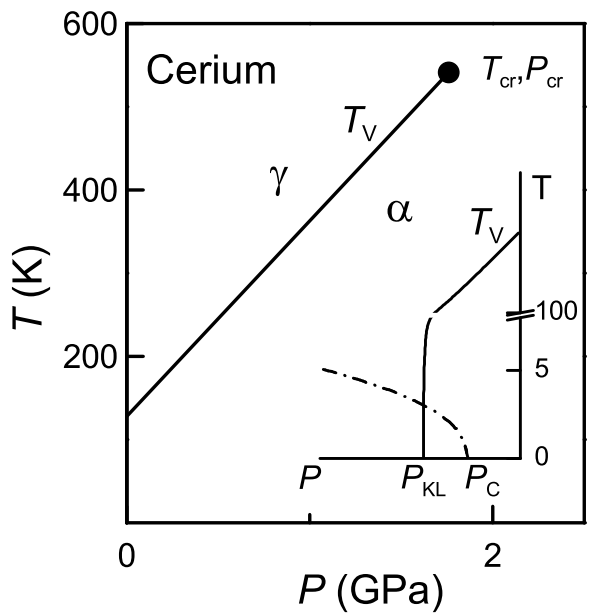}
\end{center}
\caption{Pressure-temperature phase diagram of Ce metal ($T, P$) with the first order transition $T_{\rm V} (P)$ between the $\gamma$ and the $\alpha$ phase and its critical end point. If the volume can be expanded: (i) $T_{\rm V} (P)$ will collapse abruptly at $T = 0\,{\rm K}$ for $P = P_{\rm KL}$ and (ii) magnetic ordering (dashed dotted line) will collapse at the critical pressure $P_{\rm c}$\protect\cite{Flo05}}.
\label{fig:fig1}
\end{figure}

\begin{figure}[t]
\begin{center}
\includegraphics[width=0.9 \linewidth]{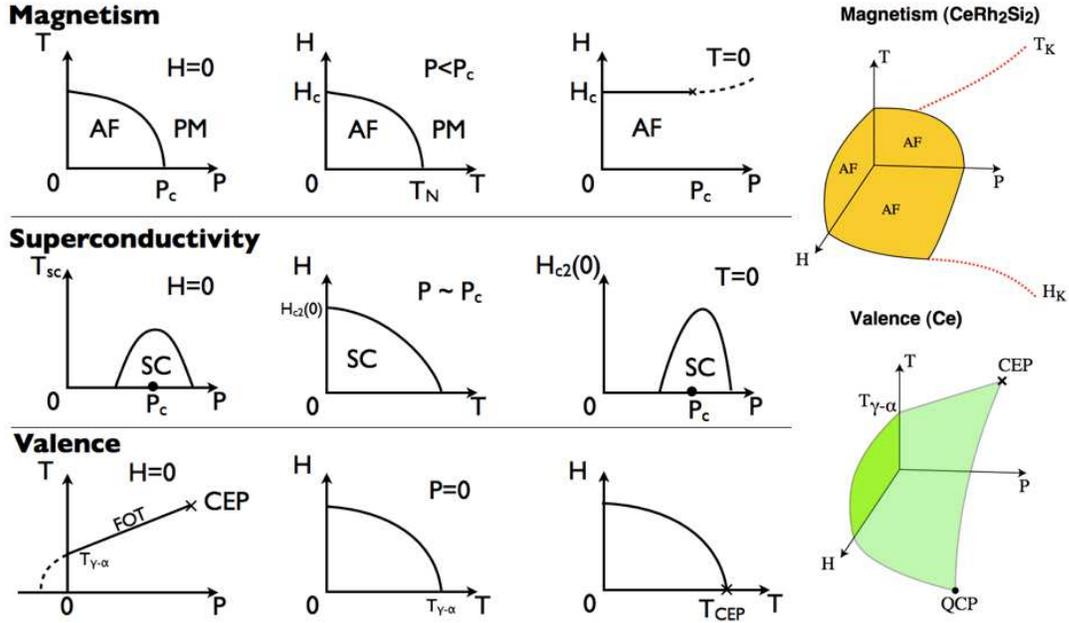}
\end{center}
\caption{Interplay between $(H, T, P)$ phase diagram for the magnetic, superconducting and valence transition. $H$ re-entrance of AF results from the interplay between magnetic and superconducting phase diagram; in Yb HFC, our guess is that there is a strong interplay between magnetic and valence phase diagram\protect\cite{Flo09}.}
\label{fig:fig2}
\end{figure}

De facto, HFC belongs to the general class of the anomalous 4f or 5f compounds
These were recognized six decades ago via the discovery of the high pressure phase diagram of Ce metal (Fig.~\ref{fig:fig1}). 
The first order transition at $T_{\gamma\alpha}$ from a PM $\gamma$ phase to a PM $\alpha$ phase is isostructural. 
It ends up at a critical end point (CEP) at $T_{\rm cr}\sim 600\,{\rm K}$ and $P_{\rm cr} \sim 2\,{\rm GPa}$. 
From high energy spectroscopy, $n_f$ in the $\gamma$ phase is near unity while even in the $\alpha$ phase $n_f$ remains high ($n_f\sim 0.9$). 
Thus the transition at $T_{\gamma\alpha}$ is connected to an abrupt change in the valence and $T_{\gamma\alpha} = T_{\rm V}$ the temperature of the valence transition. 
At ambient pressure, $T_{\rm V}(0) \sim 100\,{\rm K}$,  
such a high value prevents the establishment of long range magnetic ordering. 
However, if $T_{\gamma\alpha}$ collapses we must consider on equal footing the interplay of valence and magnetic instability~\cite{Flo05}. 
If a negative pressure could be realized, 
AF for example would appear below a N\'{e}el temperature ($T_{\rm N}$) up to a pressure $P_{\rm c}$.
An interesting problem is the consequence of an intercept between the valence transition line ($T_{\rm V}(P)$) which must collapse at some pressure $P_{\rm KL}$ at $T\to 0\,{\rm K}$ and the AF--PM boundary. 
Thus in addition to the magnetic instability at $P_{\rm c}$ 
a possible Fermi surface instability may occur. 
Furthermore, if $T_{\rm V}$ becomes comparable to any characteristic temperature of a specific bare ground state like $T_{\rm N}$ for AF, $T_{\rm Curie}$ for FM, or $T_{\rm SC}$ for SC, 
the valence transition or its fluctuations near its CEP must be considered in addition to the spin fluctuations. 
The interplay between the different phase diagram reported Fig.~\ref{fig:fig2} may lead to new coexisting matter such as the AF + SC phase of CeRhIn$_5$ discussed later. 
A discussion on valence fluctuations is given in the contribution of K. Miyake~\cite{Miy07}. 
Strikingly, a first order valence transition similar to that in Ce metal has almost never been observed in any Ce HFC. 
The main reason is that the repulsion term $U_{\rm fc}$ between the 4f electron and the light electron plays a critical role in controlling the position of the zero temperature CEP as a function of the relative $\varepsilon_f$ of the 4f level with respect to the Fermi level~\cite{Wat09}. 
Thus, very often the valence fluctuations can only be felt and the precise location of the valence quantum critical point is a ``ghost'' centered around $P = P_{\rm V}$ 
but will react to magnetic field sweeps. 
For Ce HFC, there are clear examples, e.g. CeCu$_2$Si$_2$ or CeRu$_2$Si$_2$, where $P_{\rm c}$ and $P_{\rm V}$ are well separated ($P_{\rm V} - P_{\rm c} \sim 4\,{\rm GPa}$) 
but in most cases $P_{\rm c} \sim P_{\rm V}$. 
Furthermore $P_{\rm V}$ coincides with the crossover pressure $P_{\rm CF}^\ast$
where the Kondo energy $k_{\rm B}T_{\rm K}$ surpasses the crystal field splitting $C_{\rm CF}$ and the full degeneracy $2 J + 1$ of the 4f level is recovered~\cite{Flo09}.

 Often the trivalent Yb ions with 13 4f electrons are considered as  the 4f hole analog of trivalent Ce. 
However, we want to point out two major differences~\cite{Flo09,Kne06}: 
(i) the deeper localization of the 4f electron of Yb with respect to Ce implies that the width $\Delta$ of the 4f virtual band states is one order of magnitude smaller in Yb HFC than in Ce HFC,
(ii) the larger strength of the spin-orbit coupling $\lambda$ between the $j = l - 1/2$  and $j = l + 1/2$ individual configuration of the angular momentum $l = 3$ for Yb than Ce. 
This leads to the hierarchy $\Delta > \lambda > C_{\rm CF}$ for Ce and $\lambda > \Delta \sim C_{\rm CF}$ for Yb. 
Consequently, in Ce HFC, the variation of $n_f$ is restricted between 1 to 0.84. 
Contrary, for Yb centers,  $n_f$ can vary from 0 to 1 i.\/e. with a valence changing from 2 to 3. 
These hand waiving arguments are clearly illustrated in band structure calculations (for example see in ref.\cite{Kne06} the contrast between CeRh$_2$Si$_2$  and YbRh$_2$Si$_2$). 

\begin{figure}[tbh]
\begin{center}
\includegraphics[width=0.5\linewidth]{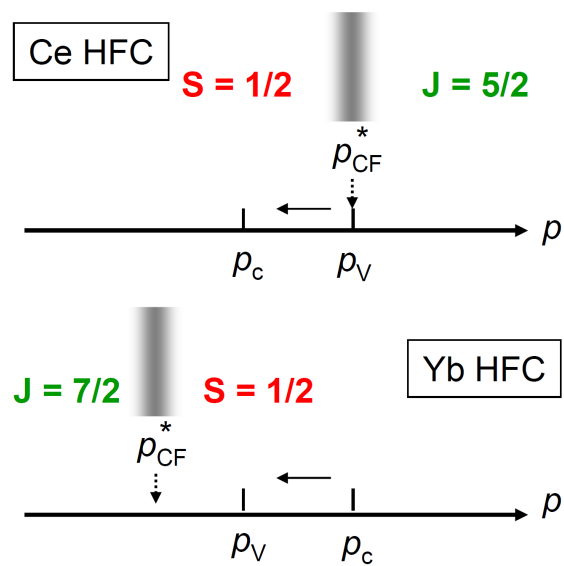}
\end{center}
\caption{Schematic difference between Ce and Yb HFC on the hierarchy between their characteristic pressure for the crystal field wipe out ($P_{\rm CF}^\ast$); for the valence transition ($P_{\rm V}$) and for the magnetic instability ($P_{\rm c}$)\protect\cite{Flo09}.}
\label{fig:fig3}
\end{figure}

\begin{figure}[b]
\begin{center}
\includegraphics[clip, width=0.4\textwidth]{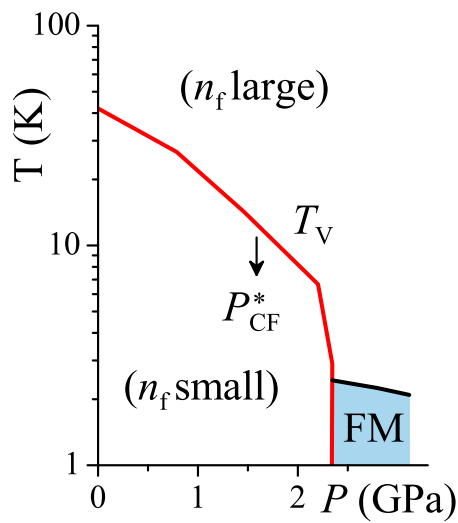}
\end{center}
\caption{Schematic ($T, P$) phase diagram of YbInCu$_4$. $T_{\rm V}$ has been determined by resistivity and NQR, $T_{\rm Curie}$ by susceptibility and NQR. $P_{\rm CF}^\ast$ indicates the crossover pressure for the mergence of crystal field effects\protect\cite{Flo09,Sar96}}
\label{fig:fig4}
\end{figure}

Another difference is that the pressure $P_{\rm CF}^\ast$ can be located deep inside the intermediate valence phase (Fig.~\ref{fig:fig3})
as pointed out two decades ago for the analysis of M\"ossbauer experiments on YbCu$_2$Si$_2$ ~\cite{Zev88} and experimentally observed for YbInCu$_4$~\cite{Sar96}. 
The latter compound represents a beautiful example of a first order valence transition with the strong interplay between valence and magnetic fluctuations (Fig.~\ref{fig:fig4}). 
In agreement with recent theoretical developments~\cite{Miy07} we want to stress that at least for Yb HFC the interplay between valence and magnetic fluctuations is strong. 
A nice illustration for this is the work on $\beta$-YbAlB$_4$~\cite{Oka10} presented by S. Nakatsuji at this conference.

\begin{figure}[tbh]
\begin{center}
\includegraphics[width=0.7 \linewidth]{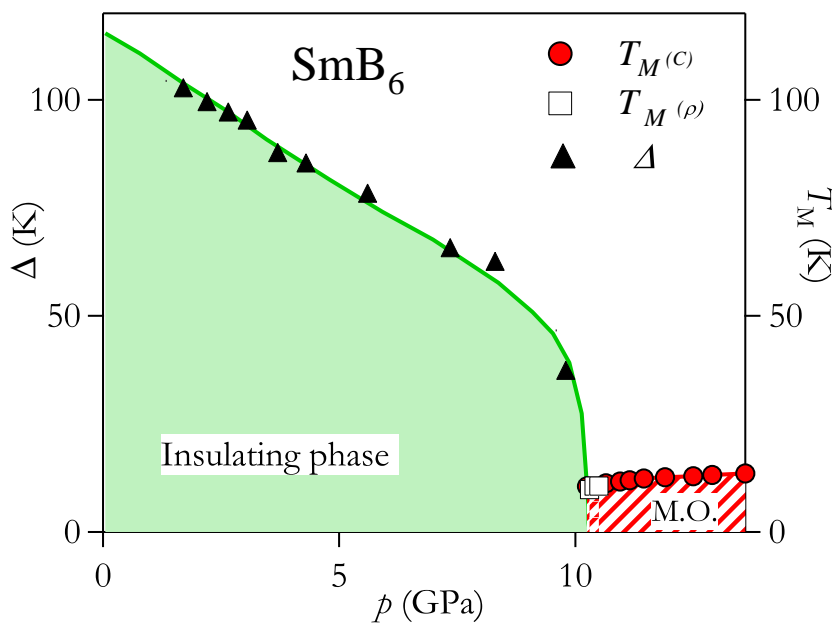}
\end{center}
\caption{(T, P) phase diagram of SmB$_6$. 
At low pressure, the ground state is PM and insulating. Above $P \sim 10\,{\rm GPa}$, 
it switches to metallic and long range magnetic ordered through a first order transition \protect\cite{Der08}.}
\label{fig:fig5}
\end{figure}

Before discussing the formation of the huge effective mass, 
let us stress that the metal-insulator transition, which can occur in anomalous rare earth compounds, remains a quite puzzling problem. 
Four decades ago,  systems like SmS~\cite{Lap81}, SmB$_6$~\cite{All79}, YbB$_{12}$~\cite{Oka98}, where the valence mixing is directly linked to the formation of an extra carrier via a relation like ${\rm Sm}^{2+}\leftrightarrow {\rm Sm}^{3+} + {\rm 5d}$,
surprisingly end up in an insulating ground state, despite the fact that the occupation number of the divalent configuration $1-n_f$ is only about 0.3.
Paradoxically, the low temperature properties seem to be renormalized to the divalent non-magnetic state even if the occupation number of the trivalent configuration $n_f = 0.7$ is higher than the divalent one. 
New sets of experiments for SmS~\cite{Bar04,Imu09} and SmB$_6$~\cite{Der08} have clearly shown  (Fig.~\ref{fig:fig5}) that 
this unconventional insulating phase is rather robust, but the ground states changes through a first order transition to a metallic long range ordered magnetic phase under pressure when $n_f$ approaches closely  to unity. 
We believe that these new data require new theoretical insights as well as a better understanding on their differences with the examples of so-called Kondo insulators such as CeNiSn~\cite{Tak90} and CeRu$_4$Sn$_6$~\cite{Pas10}.

\section{Origin of heavy fermion quasiparticle}
The huge effective mass $m^\ast \sim 100\,m_0$ is the result of two additional effects: 
(i) the renormalization of the bare band mass $m_{\rm B}$ and 
(ii) an additional contribution $m^{\ast\ast}$ due to the development of magnetic or valence correlation. 
Thus $m^\ast = m_{\rm B} + m^{\ast\ast}$. 
The last term $m^{\ast\ast}$ is directly linked to the possible occurrence of unconventional SC.
In the strong coupling theory of SC the coupling constant $\lambda$ is equal to $m^{\ast\ast}/m_{\rm B}$. 
However, fits to the temperature dependence of the superconducting upper critical field $H_{\rm c2} (T)$ indicates that $\lambda$ usually never reaches a very high value and even near a magnetic quantum critical point ($1 < \lambda < 2$)~\cite{Kne08}. 
Thus, a large part of the effective mass enhancement comes from the renormalization of the band mass $m_{\rm B}$.

\begin{figure}[tbh]
\begin{center}
\includegraphics[clip,width=0.7 \linewidth]{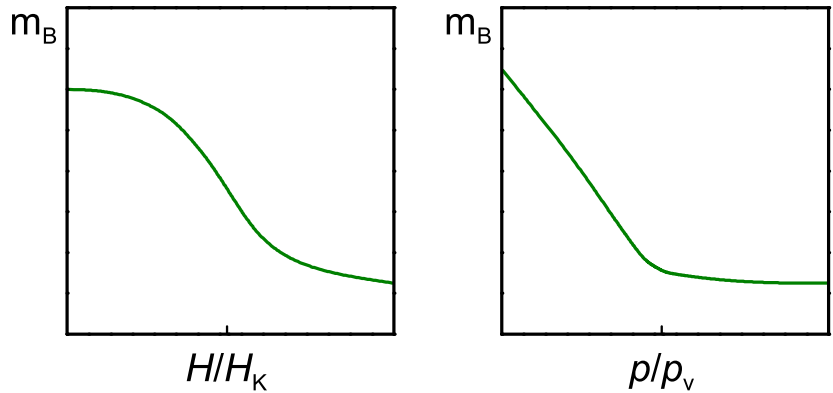}
\end{center}
\caption{Schematic representations of the field and pressure variation of the regular renormalized band effective mass.}
\label{fig:fig6}
\end{figure}
\begin{figure}[b]
\begin{center}
\includegraphics[clip,width=0.7 \linewidth]{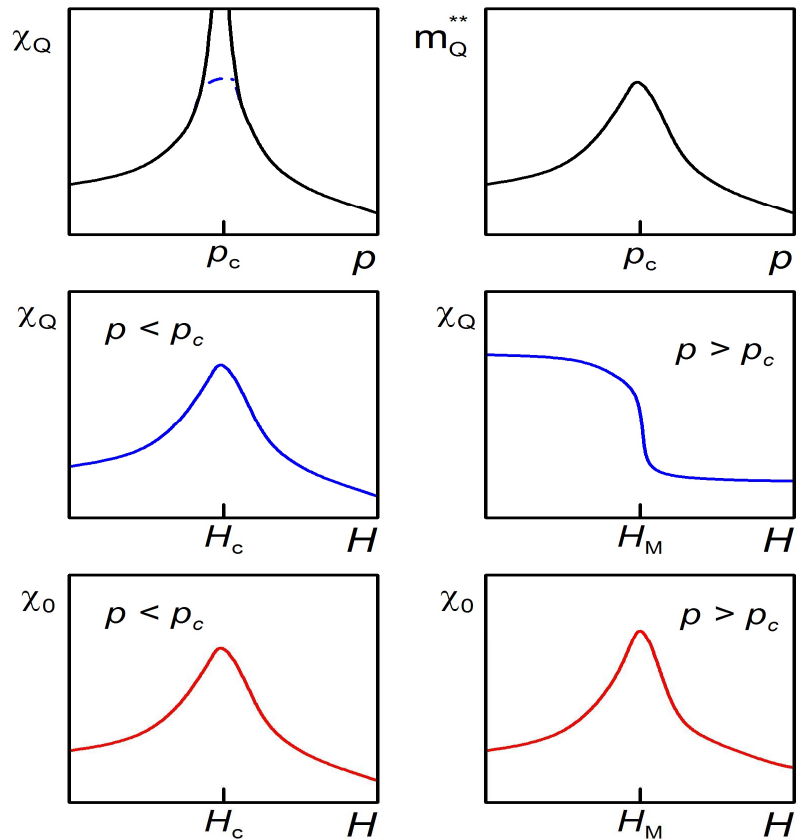}
\end{center}
\caption{$H$ and $P$ variation of the two singular (AF and FM) part contribution of $m^\ast$: $m_Q^{\ast\ast}$ and $m_0^\ast$.}
\label{fig:fig7}
\end{figure}

Figure~\ref{fig:fig6} mimics the $H$ and $P$ variation of the regular part ($m_{\rm B}$) in a scheme where the renormalized band mass $m_{\rm B}$ will be mainly given by local Kondo fluctuations. 
Thus $m_{\rm B}$ will depend on the strength of the Zeeman energy in comparison to the Kondo energy 
leading to a crossover field $H_{\rm K} \sim k_{\rm B}T_{\rm K}/g\mu_{\rm B}$ 
and on the proximity to the valence transition ($P_{\rm V}$). 
Magnetic fields can induce rapidly a strong magnetic polarization 
(often $10\,\%$ in a field of $10\,{\rm T}$) 
that can decouple spin up and spin down sub-bands with different Fermi wave vector ($k_{\rm F\uparrow}$, $k_{\rm F\downarrow}$) and effective masses. 
In the frame of global criticality, it is assumed that the regular part $m_{\rm B}$ has a weaker $H$ and $P$ dependence  than the singular part $m^{\ast\ast}$ close to the quantum singularity.

\begin{figure}[tbh]
\begin{center}
\includegraphics[width=0.7 \linewidth]{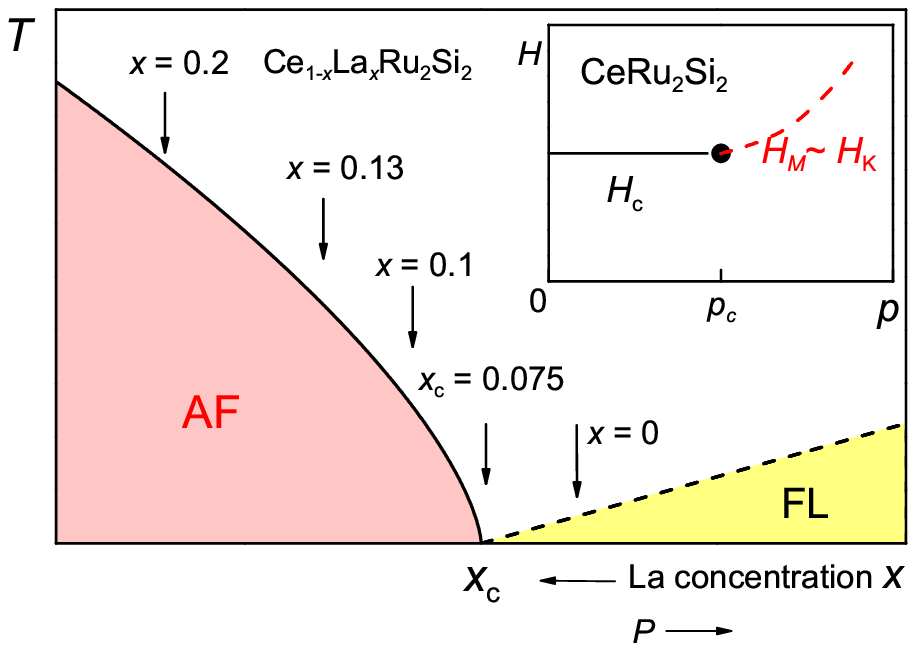}
\end{center}
\caption{Magnetic phase diagram of Ce$_{1-x}$La$_x$Ru$_2$Si$_2$ as function of the La concentration $x$. The inset shows for CeRu$_2$Si$_2$ the evolution of the metamagnetic field $H_c$ as function of pressure. $H_c$ ends at the metamagnetic critical field end points $H_{\rm c}^\ast$.
Just above the quantum  critical point, the metamagnetic phenomena is replaced by a sharp pseudo-metamagnetic crossover at $H_{\rm M}$\protect\cite{Flo05}.}
\label{fig:fig8}
\end{figure}
\begin{figure}[tbh]
\begin{center}
\includegraphics[width=0.7 \linewidth]{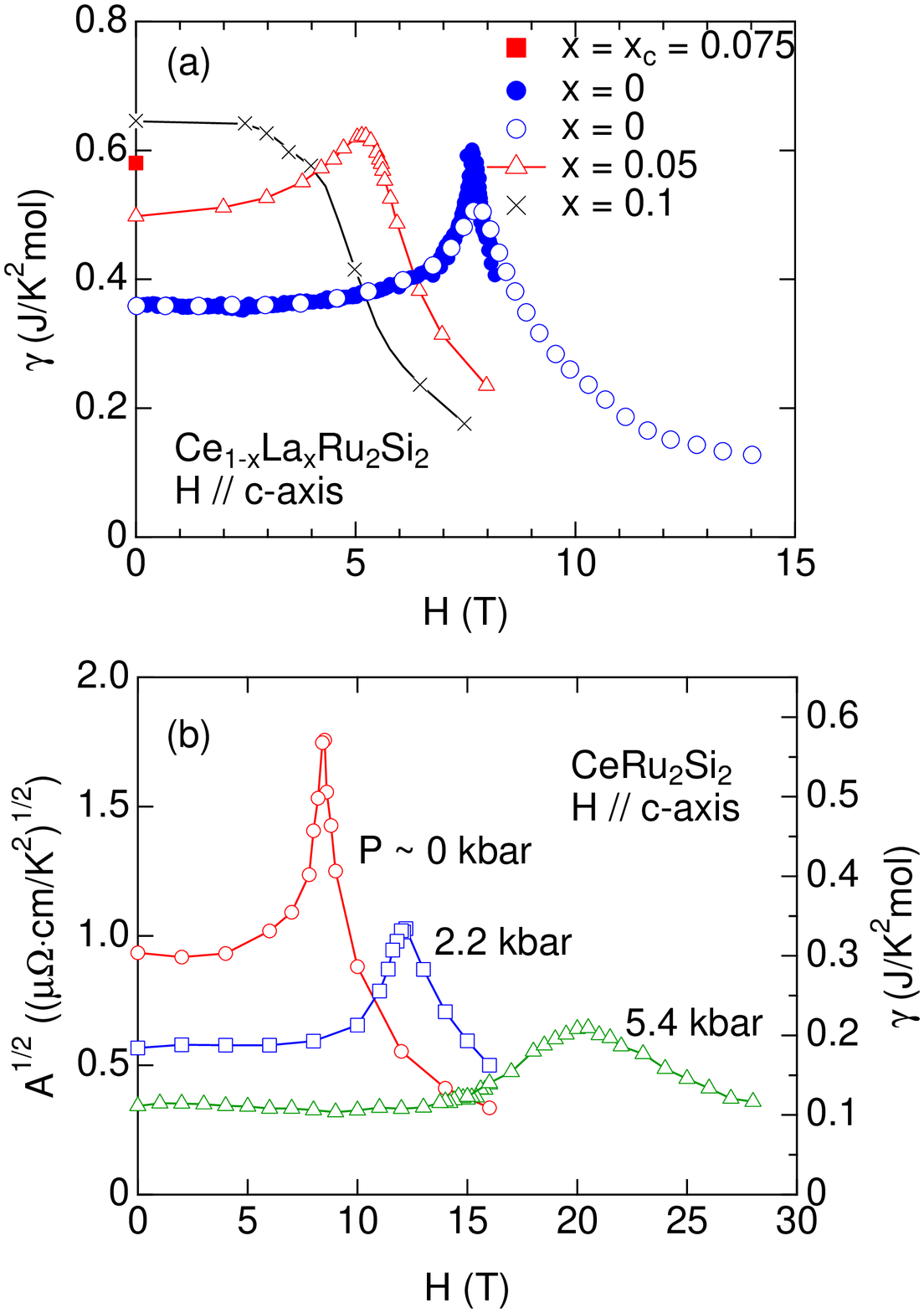}
\end{center}
\caption{(a) Field variation of the Sommerfeld coefficient of Ce$_{1-x}$La$_x$Ru$_2$Si$_2$ for $x = 0$, $x = 0.05$ which is on the AF side ($x < x_{\rm c}\sim 0.075$), and $x = 0.1$ which are located on the paramagnetic side. 
At $H_{\rm c}$ or $H_{\rm M}$ close to $P_{\rm c}$, $\gamma (P_{\rm c})$ i.~e. the value at $x_{\rm c}$ is almost the values at $H_{\rm c}$ or $H_{\rm M}$ ($\gamma (H_{\rm c} \,\mbox{or}\, H_{\rm M}$)) \protect\cite{Flo10}.
Lower panel (b) shows the variation of $\gamma (H)$ at ambient pressure compared to the field dependence of $\sqrt A$  under pressure recently deduced from resistivity experiments \protect\cite{Mat_pub}.}
\label{fig:fig9}
\end{figure}

Figure~\ref{fig:fig7} represents the qualitative $P$ and $H$ dependence of the dynamical susceptibility of an AF system at its hot spot at $Q$ 
and the corresponding variation of $m^{\ast\ast}_Q$ of an antiferromagnet 
which goes from an AF to PM phase through a metamagnetic transition. 
As shown below for CeRu$_2$Si$_2$, for $P > P_{\rm c}$ 
the metamagnetic field $H_{\rm c}$ is replaced by pseudo-metamagnetic phenomena at $H_{\rm M}$  
which can be considered as a continuation of the metamagnetic critical end point $H_{\rm c}^\ast$. 
Furthermore, the induced magnetization leads to the development of FM fluctuations that originate from the AF correlations with enhanced fluctuations at $H_{\rm c}$ or $H_{\rm M}$.
Hence an extra contribution of $m^\ast$ ($m^{\ast\ast}_Q$) will appear in magnetic field. 
It has been recently demonstrated that for AF systems both singular contributions $m^{\ast\ast}_Q$ and $m_0^{\ast\ast}$ are comparable close to $P_{\rm c}$~\cite{Mis09}.

Before discussing in more detail tricriticality phenomena for two selected examples, the CeRu$_2$Si$_2$ series (AF) and of UGe$_2$ (FM), 
we point out that there is no consensus about the possibility of a divergence of $m^\ast$ in HFC. 
For the global scenario often referred to as the conventional spin-fluctuation approach~\cite{Mor95}, 
there is no divergence of $m^{\ast\ast}_Q$ at the AF singularity, even for a second order quantum critical point.
For FM material a possible divergence of $m^{\ast\ast}_0$ at the FM quantum critical point 
is totally wiped out by the first-order nature of the singularity~\cite{Bel05_RMP}. 
In the local quantum criticality scenario, 
a divergence of effective mass is often invoked  at its magnetic quantum critical field notably for YbRh$_2$Si$_2$ $H_{\rm C}$~\cite{Geg08,Fri09}.
We consider there is no evidence of such a divergence~\cite{Kne06}. 
Effectively a maximum of $m^\ast$ occurs at $H_{\rm c}$ with a very sharp decrease of $m^\ast$ for $H>H_{\rm c}$. This situation is 
quite comparable to that of the polarized phase of CeRu$_2$Si$_2$ which is discussed hereafter  
and in contrast with a smooth $H$ increase of $m^\ast$ going from $H = 0$ to $H = H_{\rm c}$. 
It is obvious that a new set of experiments is required to settle this point. 
Possible candidates to study are cage compounds like the recently investigated YbT$_2$Zn$_{20}$ systems~\cite{Jia07,Hon10}. In these system a  $\gamma$ term 
as large as $10\,{\rm J\,mole^{-1}K^{-2}}$ has been reported with strong magnetic field effects and they open new possibilities to study the formation of the effective mass.

\section{Tricriticality: the AF CeRu$_2$Si$_2$ series --- the FM UGe$_2$  case}

CeRu$_2$Si$_2$ has been highly studied as it is located close to the AF instability on the PM side  ($P_{\rm c} \sim -0.3\,{\rm GPa}$). 
It is also one of the few HFC where the Fermi surface is fully determined~\cite{Suz10} 
and thus allows a real proof on the itinerant 4f electron description. 
Furthermore, extensive inelastic neutron experiments show that 
the AF correlations develop up to $T_{\rm corr} = 60\,{\rm K}$ 
i. e. at temperature higher than the single Kondo temperature $T_{\rm K} \sim 25\,{\rm K}$~\cite{Ros88}. 
Clearly, there is no evidence of an intermediate temperature regime 
where the effective mass enhancement will be dominated by $m_{\rm B}$. 
This is well demonstrated by the continuous increase of the electronic Gr\"{u}neisen parameter ($\Omega (T)$) on cooling~\cite{Lac89,Wei10} 
since if, in a finite $T$ range, the free energy will be controlled by a main parameter $T^\ast$, 
a constant Gr\"{u}neisen regime must emerge. 
The large value of the ratio $T_{\rm corr}/ T_{\rm K}$ agrees with the conclusion that 
in HFC the 4f centers behave as single site Kondo centers only at high temperatures~\cite{Yan10}.

\begin{figure}[tbh]
\begin{center}
\includegraphics[width=0.9 \linewidth]{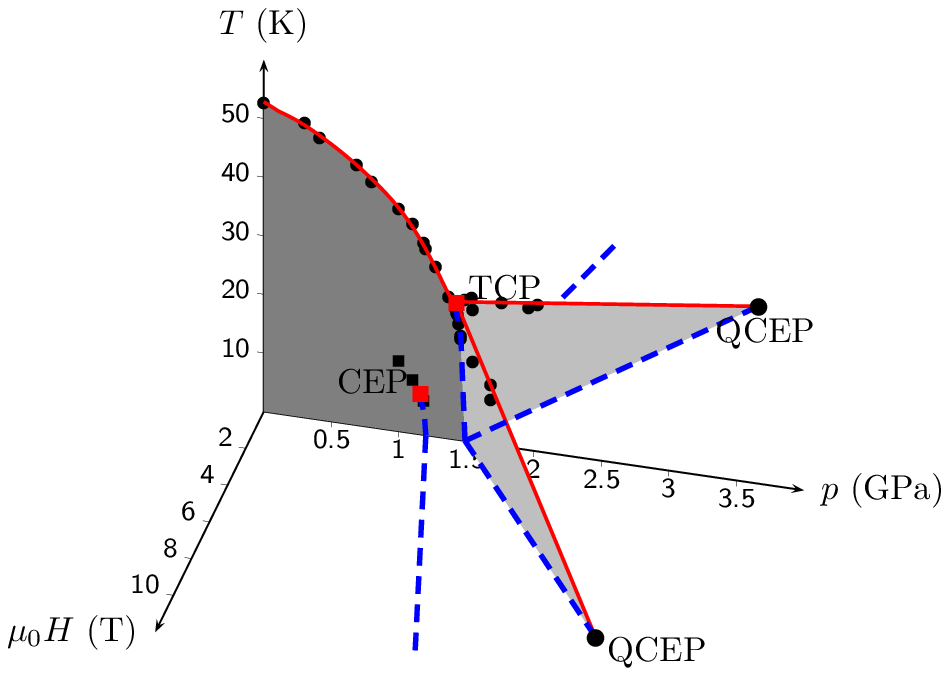}
\end{center}
\caption{UGe$_2$ wings with the tricritical pressure $P_{\rm c}$ at $T = 0\,{\rm K}$ and the collapse of the first order metamagnetic transition at $T = 0$ for $P_{\rm QCP} > 3\,{\rm GPa}$ and $H_{\rm QCP} > 16\,{\rm T}$\protect\cite{Tau_pub}}
\label{fig:fig10}
\end{figure}

Expanding the lattice volume of the CeRu$_2$Si$_2$  by La doping, induces 
AF order above a critical doping level ($x_{\rm c}$) near $7\,\%$ La (Fig.~\ref{fig:fig8}). 
The Ising character of the localized spin leads to clear first order metamagnetic phenomena at a field $H_{\rm c}\sim 4\,{\rm T}$. 
The value of $H_{\rm c}$ is weakly pressure dependent leading to a critical end point at $\sim 4\,{\rm T}$ for $P = P_{\rm c}$ on $x=x_{\rm c}$.
Entering in the PM regime above $P_{\rm c}$, the metamagnetic transition ($P < P_{\rm c}$, $H = H_{\rm c}$) changes to sharp pseudo-metamagnetic crossover at a field $H_{\rm M}$. Furthermore the characteristic field $H_{\rm c}$, or $H_{\rm M}$, corresponds to a critical value $M_{\rm c}\sim 0.4\,\mu_{\rm B}$ of the magnetization. 
This points out that the increase of the volume of the majority spin band is the source of the metamagnetic and pseudo-metamagnetic phenomena. 
Microscopic inelastic neutron scattering experiments show~\cite{Flo04,MSat01} that, for CeRu$_2$Si$_2$,  a
drastic softening of the FM fluctuation occurs at $H_{\rm M}$ 
while the intensity of the AF correlation collapses at $H_{\rm M}$ with almost a persistence of the same energy for the AF fluctuation itself~\cite{Ray98}. 
It was recently stressed~\cite{Mat_pub,Flo10} that close to $P_{\rm c}$
a quasi-convergence  between the effective mass $m^\ast$ ($H_{\rm c}$ or $H_{\rm M}$) at $H_{\rm c}$ or $H_{\rm M}$ 
and the effective mass $m^\ast$($P_{\rm c}$ ) at $P_{\rm c}$ in zero magnetic field exists: 
$m^\ast(P_{\rm c}) \sim m^\ast(H_{\rm c})$ as
illustrated in Fig.~\ref{fig:fig9}. 
For $x = 0.1$ with a N\'{e}el temperature $T_{\rm N} \approx4\,{\rm K}$ the Sommerfeld coefficient at $H=0$ is about $\gamma = 0.65\,{\rm J\,mole^{-1}K^{-2}}$ and decreases monotonously with increasing field.  
At the critical point $x_{\rm c} \sim 0.075$ we find $\gamma \approx 0.6\,{\rm J\,mole^{-1}K^{-2}}$, 
quite similar to value for CeRu$_2$Si$_2$ at $H_{\rm M}$. 
Escaping further from $P_{\rm c}$ by applying a pressure on CeRu$_2$Si$_2$, 
 the $\gamma$ term at $H = 0$ decreases strongly in good agreement with the large magnitude of the Gr\"{u}neisen parameter $\Omega (0) \sim +200$. The relative enhancement of $\gamma (H_{\rm M})/\gamma (H = 0)$ is at least at low pressure weakly pressure dependent~\cite{Mat_pub,Flo10}. 
Under pressure, correlations are still efficient but form a quite smoother pseudogap 
than the one built at $P = 0$; 
the FM interaction is created as the spin up and spin down sub-bands decouple 
when a critical value of the magnetization is reached~\cite{Flo05}. 
Approaching the valence fluctuation regime, near $P = 4 \,{\rm GPa}$, 
such a picture will certainly fail. 
It is worthwhile to point out that in the research on quantum criticality in HFC, 
most of the results are concentrated on the PM side ($P > P_{\rm c}$  for Ce HFC); 
in Grenoble attempts are made now to look more carefully to the AF side ($P < P_{\rm c}$) and thus to make quantitative comparisons with the recent developments on tricriticality.

It is well known that a FM state can be induced by magnetic field at $H_{\rm c}$ through a first order transition from a PM ground state at a pressure $P$ higher than the critical pressure $P_{\rm c}$ of the FM--PM instability at $H = 0$~\cite{Shi64}. 
A nice example is UGe$_2$~\cite{Hux01} as, at the first order FM critical pressure $P_{\rm c}\sim 1.49\,{\rm GPa}$, 
the sublattice magnetization jumps from the FM phase to the PM one from $M_0=0.9\,\mu_{\rm B}$ to zero. 
In zero field, the transition line $T_{\rm Curie} (P)$ between FM and PM changes from second order to first order at a tricriticality point $T_{\rm TCP} \sim 24\,{\rm K}$, $P_{\rm TCP}\sim 1.42\,{\rm GPa}$~\cite{Tau_pub,Kab10}. 
The domain of the first order transition at $H = 0$ is quite narrow $(P_{\rm c} - P_{\rm TCP})/P_{\rm TCP}\sim 0.05$, 
but recent resistivity measurements~\cite{Tau_pub} as well as thermal expansions~\cite{Kab10} data show that FM wings~\cite{Bel99} appear under magnetic field, 
and a first order transition will occur. They will only collapse at $T = 0\,{\rm K}$ for $P > P_{\rm QCEP} \sim 3\, {\rm GPa}$ i.~e. $2 P_{\rm c}$ around $H_{\rm QCEP} \sim 10\,{\rm T}$. 
For UGe$_2$, a large pressure is needed ($2 P_{\rm c}$) to reach a PM ground state at any magnetic field (Fig.~\ref{fig:fig10}). 
When the size of the FM sublattice magnetization close to $P_{\rm c}$ decreases, 
as for UCoGe ($M_0 \sim 0.07\,\mu_{\rm B}$)~\cite{Huy07}, 
no detection of FM wings has been reported. 
Thus the large separation between $P_{\rm QCEP}$ and $P_{\rm c}$ in UGe$_2$ is clearly linked to the large jump of $M_0$ at $P_{\rm c}$. Due to the insertion of Yb ions in a cage, one may suspect a very high value of $m_B$ which will furthermore highly vary with the magnetic field.


\section{Evidence of $m_{\rm B}$ and $m^{\ast\ast}$ in FM superconductors}

Three U based FM-SC have been discovered, namely UGe$_2$~\cite{Sax00}, URhGe~\cite{Aok01}, and UCoGe~\cite{Huy07}. 
The respective values of $M_0$ at $P = 0$ are $1.4\,\mu_{\rm B}$, $0.4\,\mu_{\rm B}$ and $0.07\,\mu_{\rm B}$ and their Curie temperatures $T_{\rm Curie} = 50\,{\rm K}$, $9.5\,{\rm K}$, and $2.8\,{\rm K}$. 
In UGe$_2$, SC appears only under pressure with a maximum of the critical SC temperature 
$T_{\rm SC, max}\sim 0.7\,{\rm K}$ reached close to the pressure $P_{\rm x}$ where the system switches one the ferromagnetic FM2 phase to another ferromagnetic FM1 phase~\cite{Hux01}. 
The discoveries of SC in URhGe ($T_{\rm SC}\sim 0.25\,{\rm K}$) and in UCoGe ($T_{\rm SC} \sim 0.7\,{\rm K}$) already at $P = 0$ open the possibility of careful measurements even at ambient  pressure. 
UGe$_2$ and UCoGe have a PM ground state above respectively $P_{\rm c}\sim 1.49\,{\rm GPa}$~\cite{Hux01} 
and $P_{\rm c}\sim 1.2\,{\rm GPa}$~\cite{Has08_UCoGe,Has10,Slo09},
while in URhGe an initial $P$ increase of $T_{\rm Curie}$ is observed, at least up to $P = 13\,{\rm GPa}$~\cite{Har05_pressure}. 
%
\begin{figure}[tbh]
\begin{center}
\includegraphics[width=0.7 \linewidth]{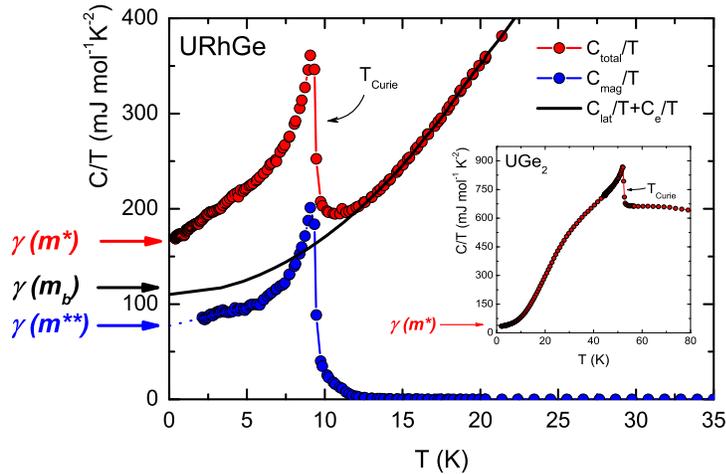}
\end{center}
\caption{Evidence of the $m_{\rm B}$ strength at ambient pressure for URhGe from the temperature dependence of $C/T$. The inset shows $C/T (T)$ for UGe$_2$.}
\label{fig:fig11}
\end{figure}
%
In contrast to UGe$_2$, where SC exists only in the FM side, in UCoGe, SC persists also in the PM regime above $P_{\rm c}$. 
SC in URhGe is suppressed above $p \approx 4\,{\rm GPa}$~\cite{Miy09}. 
For the low $T_{\rm Curie}$ materials
(URhGe~\cite{Har_pub} and UCoGe~\cite{Aok_pub}), 
the measurements of the specific heat (see Fig.~\ref{fig:fig11} for URhGe) show clearly the occurrence of a term linear in temperature above $T_{\rm Curie}$ which can be associated to the formation of the renormalized band mass $m_{\rm B}$ already at $T_{\rm Curie}$. 
Furthermore applying a magnetic field along the easy $c$ axis of magnetization leads to quench the FM fluctuation (thus to the collapse of $m_0^{\ast\ast}$) 
and to restore the sole $m_{\rm B}$ contribution. 
By contrast, in the high $T_{\rm Curie}$ ($\sim 50\,{\rm K}$) material as UGe$_2$~\cite{Har09_UGe2}
there is no mark of renormalized band mass contribution to the specific heat above $T_{\rm Curie}$ at $P = 0$ as its characteristic temperature $T_{\rm B}$ may be quite similar than $T_{\rm Curie}$. 
The ''hidden'' character of $m_{\rm B}$ is quite similar to the previous discussion made for CeRu$_2$Si$_2$.

\section{Link between $m^{\ast\ast}$ and $T_{SC}$ in the ferromagnetic superconductor URhGe}

\begin{figure}[b]
\begin{center}
\includegraphics[width=0.5 \linewidth]{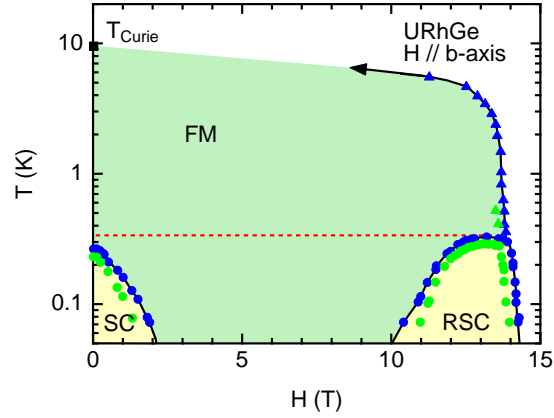}
\end{center}
\caption{$H, T$ phase diagram of URhGe at ambient pressure with the occurrence of low field SC and $H$ reentrant superconductivity at $H_{\rm R}$ which is directly associated with the field mass enhancement (see \protect\cite{Lev05,Miy08}).}
\label{fig:fig12}
\end{figure}

\begin{figure}[tbh]
\begin{center}
\includegraphics[width=0.7 \linewidth]{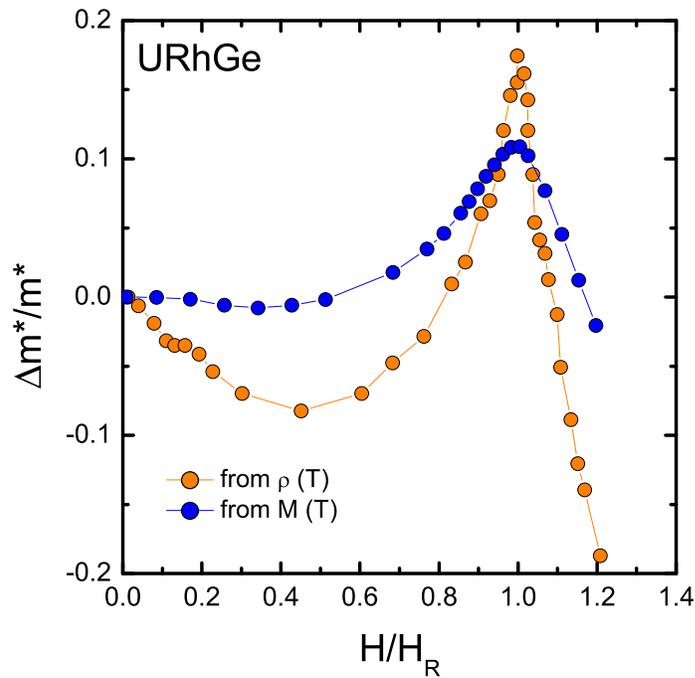}
\end{center}
\caption{$H$ enhancement of the effective mass as deduced from the $A$ coefficient of the resistivity measurements down to $80\,{\rm mK}$ and from the temperature dependence of magnetization measurements down $1.5\,{\rm K}$ using the Maxwell relation \protect\cite{Har_pub}.}
\label{fig:fig13}
\end{figure}

URhGe is a very nice example to illustrate the link between the strength of $m^{\ast\ast}$ and its feedback to SC. 
A orifinally discovery was that the  application of a magnetic field along the $b$ hard-axis induces the re-orientation of the magnetization from the easy $c$ axis to the $b$ axis at a field $H_{\rm R} \sim 12\,{\rm T}$ and this re-orientation is directly  associated with a $H$ re-entrance of SC (Fig.~\ref{fig:fig12})~\cite{Lev05}. 
To verify that the driving phenomena is an enhancement of $m^{\ast\ast}$, 
careful resistivity experiments as a function of pressure and magnetic field have been realized~\cite{Miy09,Miy08}. 
Furthermore, this coupling was verified by detailed magnetization ($M$) studies. 
Figure~\ref{fig:fig13} compares the field enhancement of $\Delta m^\ast/m^\ast$ detected by resistivity ($\rho$) and magnetization~\cite{Har_pub2}. 
Taking into account that magnetization measurements are limited down to  $T=1.5$~K while resistivity has been realized down to $70\,mK$,
the agreement can be considered  excellent. 
Assuming that the $H$ enhancement of $m^\ast$ is that of $m^{\ast\ast}_0$, 
via the McMillan-type formula $T_{\rm SC} (m^\ast) = T_0 \exp (-m^\ast/m^{\ast\ast})$, 
the field re-entrance (RCS) of SC as well as its collapse at a pressure $P_{\rm RSC} \sim 2\,{\rm GPa}$, which is two times smaller than the pressure $P_{\rm LFSC} \sim 4\,{\rm GPa}$ where SC would collapse in a low field, were quantitatively explained. 
To our knowledge, it is the first case where a link between the mass enhancement and the appearance of SC has been shown successfully on a given material between the ($P, H$) variation of $m^{\ast\ast}$ and of $T_{\rm SC}(m^{\ast\ast})$.

An anomalous temperature dependence of the upper critical field $H_{c2} (T)$ has also been reported in UCoGe when the field is applied perpendicular to the easy $c$ axis \cite{Aok2009}. The common point with URhGe is that in this Ising type ferromagnet when $H$ is applied along the hard magnetization axis (see \cite{Har_pub2}) the Curie temperature decreases with $H$, as proofed theoretically \cite{Min10}, which will obviously lead to field mass enhancement. In UCoGe, there is no evidence that the anomalies of $H_{c2}$ for field perpendicular to the $c$ axis is correlated with a spin re-orientation. The key parameter may be the size of the field induced moment for example $\chi_bH$ ($\chi_b$ is the susceptibility along the hard $b$ axis) by comparison to $M_0$. It will drive the spin re-orientation if $\chi_b$ is greater than $\chi_c$ as it is the case for URhGe. Any induced supplementary component along the $c$ axis via an angular misalignment of the $b$ axis versus the magnetic field direction will wipe out rapidly the phenomena as an extra magnetic component along $c$ leads to the opposite to a decrease of the strength of the spin fluctuations. Recently a theory on field induced Ising spin fluctuations has been developed \cite{Tad_pub}.

\section{Interplay of antiferromagnetism and superconductivity: before and after the discovery of 115 compounds}

\begin{figure}[b]

\begin{center}
\begin{minipage}{0.49\hsize}
\begin{center}
\includegraphics[width=.9\hsize,clip]{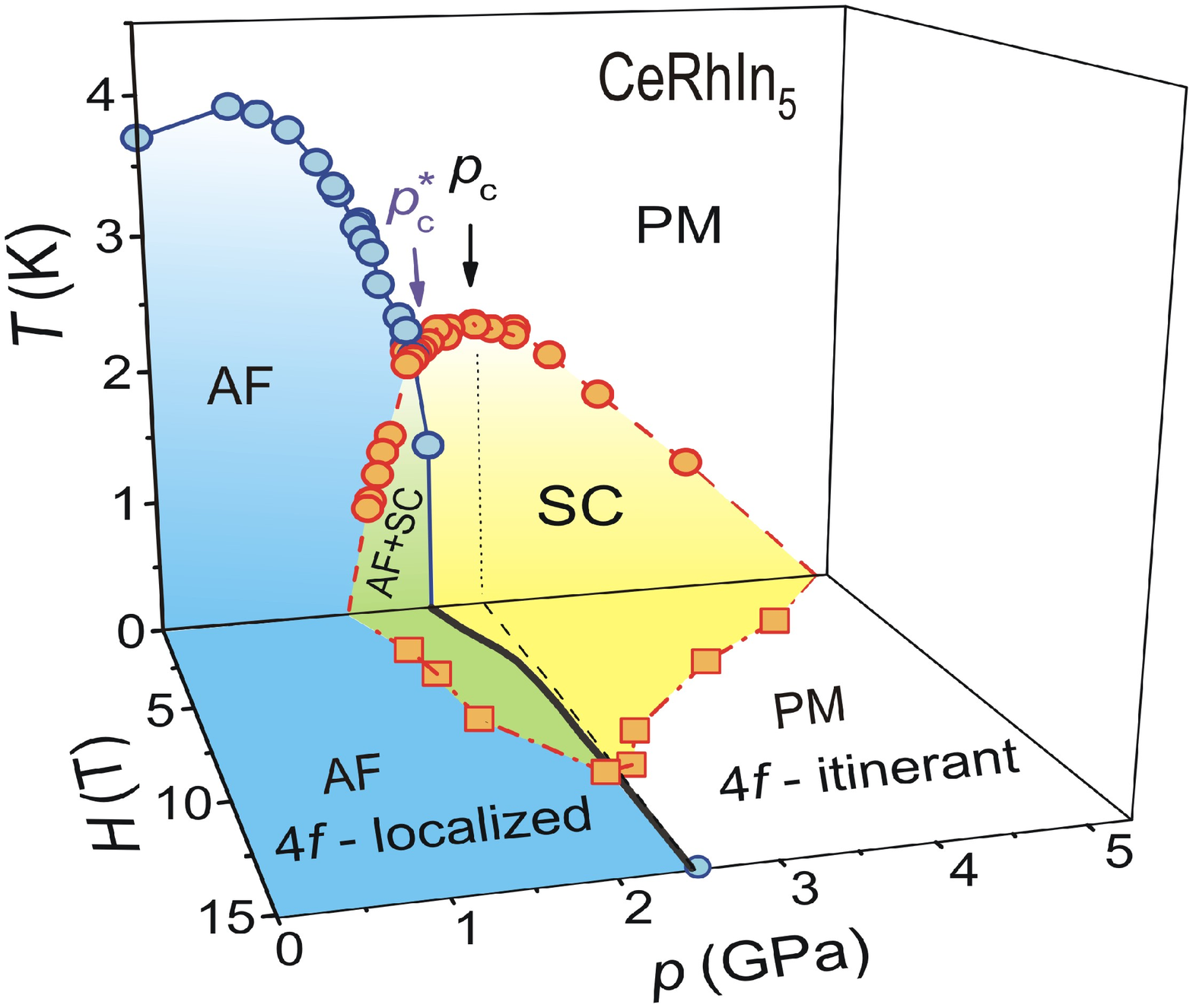}
\end{center} 
\end{minipage}
\begin{minipage}{0.49\hsize}
\begin{center}
\includegraphics[width=.8\hsize,clip]{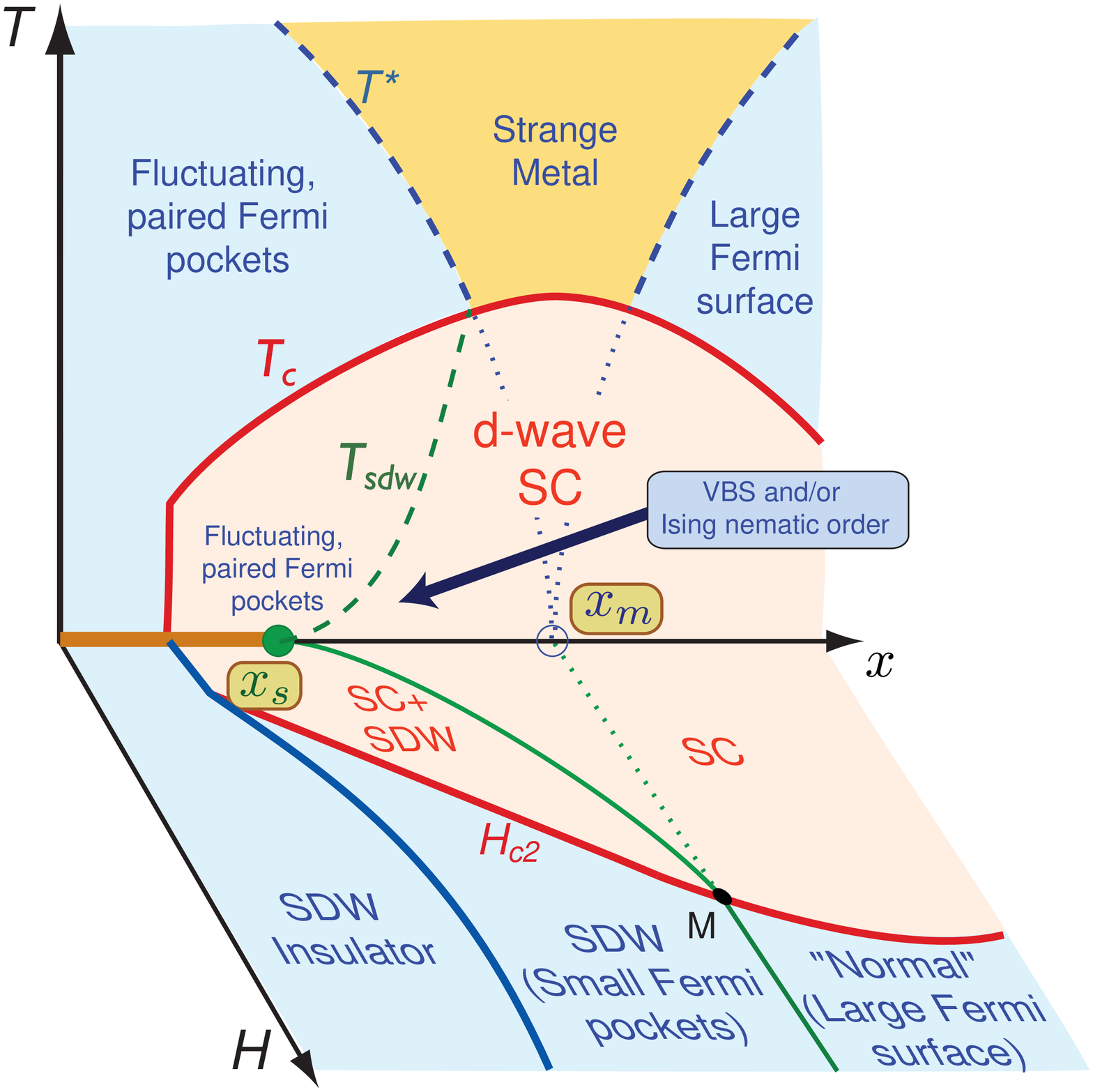}
\end{center} 
\end{minipage}
\end{center}
\caption{($H, T, P$) phase diagram of CeRhIn$_5$ compared to the one proposed for the high $T_{\rm c}$ superconductor  \protect\cite{Kne09,Sac09}.}
\label{fig:fig14}
\end{figure}

Careful studies on the interplay between AF and SC were quite limited before the appearance of the Ce-115 HFC as  the SC dome near $P_{\rm c}$  occurs at low temperature ($T = 0.6\,{\rm K}$ or lower) far below the maximum of the N\'{e}el temperature ($T_{\rm N, max}$) of the AF transition. 
The discovery of SC in Ce-115 compounds (discussed by T. Parks at this conference) opens new opportunities to study in detaile the interplay of AF and SC, notably on the CeRhIn$_5$, via resistivity, microcalorimetry, and neutron scattering as both transition temperatures are comparable. 
Figure~\ref{fig:fig14} represents the ($T, P, H$) phase diagram of CeRhIn$_5$ determined in Grenoble~\cite{Kne06,Kne04}.

The main point is that at $H = 0$, above $P=P_c^\ast \approx 2\,{\rm GPa}$
 where $T_{\rm SC} = T_{\rm N} (P=P_{\rm c}^\ast)$, 
the opening of a large SC gap leads to the rapid suppression of antiferromagnetism at $P=P_{\rm c}^\ast + \varepsilon$. However, in absence of SC, AF will survive up to $P_{\rm c} \sim 2.5\,{\rm GPa}$. From macroscopic measurements \cite{Kne06,Kne04,Par06}, the coexisting domain of AF + SC is small with the onset of SC at $P_{\rm S}^{+}\sim 1.5\,{\rm GPa}$ and a disappearance of AF close to $P_{\rm c}^\ast$. 
Looking only to the specific heat anomaly at the phase transition from the AF phase to the AF + SC state indicates that 
the transition is not at all BCS like. 
NMR experiment give two piece of evidence~\cite{Yashima07} that (i) the ground state appear homogeneous i.~e. with no phase separation between AF and SC components, (ii) the magnetic structure switches from incommensurate to commensurate.

The striking new phenomena in magnetic field is the lowering of the strength of SC and the creation of vortices. 
Both phenomena lead to the $H$ re-entrance of AF to occur up to the pressure $P_{\rm c}$~\cite{Kne06,Kne04,Par06}. 
Up to now, no microscopic studies have been realized in the high magnetic field AF + SC phase. 
As indicated on Fig.~\ref{fig:fig14}, two clear limits exist: 
(i) the low pressure $P<P_{\rm s}^{+} = 1.5\,{\rm GPa}$ regime,
where only AF occurs, the Fermi surface corresponds to a local behavior of the 4f electron, 
(ii) the high pressure limit with $P>P_{\rm c}$, where a pure SC phase exists up to $H_{\rm c2}$ and the Fermi surface implies an itinerant character of the 4f electrons~\cite{Kne09}. 
Between $P_{\rm s}^+$ and $P_{\rm c}$ for $H < H_{\rm c2}$, the FS must evolve in pressure and magnetic field between the AF + SC phase and the pure SC phase.

The ($H, P, T = 0\,{\rm K}$)  phase diagram of CeRhIn$_5$ is quite analogous to that proposed for the high $T_{\rm c}$ superconductors (e.g. YBa$_2$Cu$_3$O$_{7-x}$) with the change from spin density wave insulator to SC + spin density wave with small Fermi surface pockets and then ending to a pure SC phase with ''normal'' large Fermi Surface~\cite{Sac09} as a function of the hole concentration.  
Thanks to the recent discoveries of new SC materials such as the pnictides, e.g.(see \cite{Ish09}) 
there are now many examples of an interplay between AF and SC with the occurrence of either an homogeneous phase or a phase separation. Precise experiments can be realized in future to reach a clearer view.

\begin{figure}[tbh]
\begin{center}
\includegraphics[width=0.7 \linewidth]{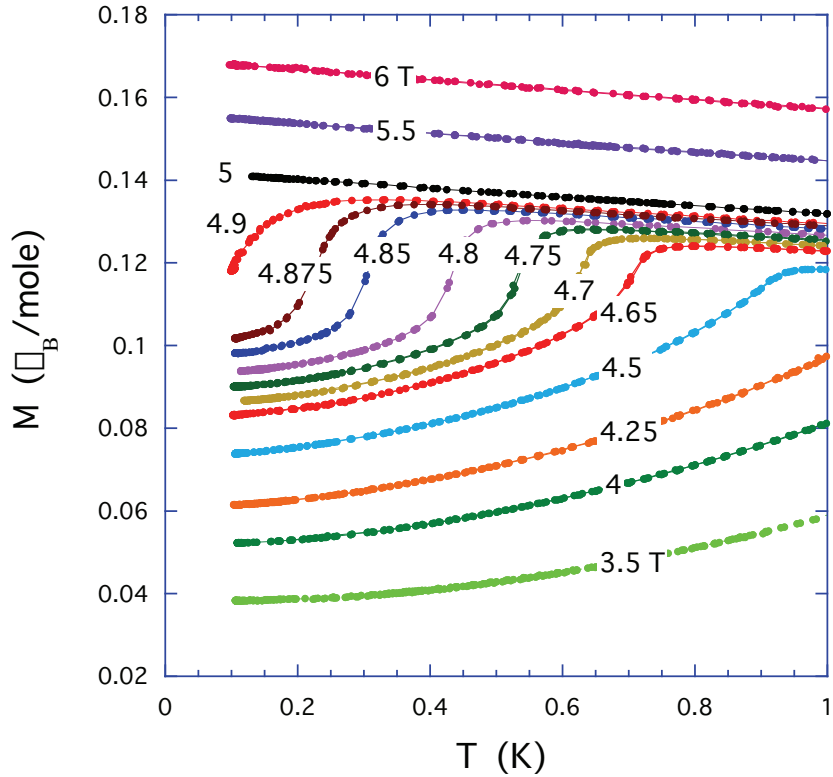}
\end{center}
\caption{Magnetization as function of temperature at constant field $H \parallel c$ of CeCoIn$_5$ showing clearly the singular magnetic behaviour right at $H_{\rm c2}$.}
\label{fig:fig15}
\end{figure}

CeCoIn$_5$ corresponds to the case of a pure SC phase at ambient pressure. 
A consequence of the previous remarks that high magnetic polarization can be reached plus the weakness of the Pauli limit of $H_{\rm c2}(0)$ with respect to the orbital one pushes the interpretation of the appearance of a new high magnetic field low temperature phase (HFLT) to be a FFLO phase~\cite{Bia02,Mat07} as predicted four decades ago by Fulde-Ferrel~\cite{Ful64} and Larkin-Ovchinnikov~\cite{Lar64}. 
However recent NMR~\cite{You07,Kou10} and neutron scattering  experiments~\cite{Ken08,Ken10} for $H \parallel a$ prove clearly that the HFLT phase has a long range magnetic component and the wave vector does not change with the direction of the magnetic field. 
Here, we will not discuss in detail the physics of the HFLT, but recently we have realized a new set of magnetization experiments with $H \parallel c$ in order to clarify  whether exactly at $H_{\rm c2}(0)$ 
a quantum singularity exists since the $H_{\rm c2}(0)$ point is not associated with a magnetic quantum criticality~\cite{Ron06}. 
Our magnetization ($M(H)$) measurements performed down to $80\,{\rm mK}$ are mainly an improvement by a fine zoom in the $(H-H_{\rm c2})/H_{\rm c2}$ window of previous published data~\cite{Tay02}. 
Figure~\ref{fig:fig15} shows at constant field value, the temperature variation of $M(H, T)$. 
This spectacular result is that, below $H_{\rm c2}(0)$, $M(T)$ approaches a $T^2$ dependence with a $T$ decrease of $M(H)$ on cooling while above $H_{\rm c2}(0)$, $M(H, T)$ does not obey a $T^2$ Fermi liquid law but a quasi linear $T$ dependence with an increase of $M(H)$ on cooling. Using Maxwell relations it can be clearly shown that coming from the SC phase a sharp maxima of the $\gamma$ term of the specific heat will occur at $H_{c2} (0)$. The extrapolation of the suspected divergence of the effective mass from the field dependence of the $A$ coefficient of the resistivity measured at fields above $H_{c2} (0)$ at $H_{QCP} \sim 4.3$ T \cite{How10} is not marked by any change of the field dependence of the $\gamma (H)$ at $H_{QCP}$. Experimentally it should be possible to push down the magnetization measurements below 20 mK where a Fermi liquid temperature dependence of $M$ may be obeyed on both sides of $H_{c2}(0)$ and to establish if at $H_{c2}(0)$ the field dependence of $\gamma$ has no discontinuity linked to the first order nature of the transition. The unusual behavior, notably the quasi linear $T$ variation of $M$ above $H_{c2}(0)$ at least down to 80 mK is a consequence of the interference between magnetic fluctuations and superconducting correlations as also discussed recently in organic and pnictide superconductors \cite{Doi10}.
Thus, there is a clear evidence that $H_{\rm c2}(0)$ is a quantum singularity caused by the interplay of normal and superconducting  properties. It was proved that AF criticality can be induced by the onset of superconductivity around $H_{c2} (0)$ \cite{Fuj08}.  The exciting phenomena in CeCoIn$_5$ is that magnetism is glued to SC as if the $H$ weakening of the SC gaps leads to a restoration of a pseudogap structure favorable to AF.  Even more the first order magnetization jump is quite similar macroscopically to the metamagnetic phenomena described previously for the CeRu$_2$Si$_2$ series. It has been proposed that the tendency of field induced AF ordering is linked to strong Pauli depairing (as can be viewed for pseudometamagnetism) and strongly favored by unconventional SC gap node along the AF modulation \cite{Ike2010}.
The localisation of a magnetic quantum critical point in the  $(H, T, P)$ phase diagram of CeCoIn$_5$ in the absence of SC is still an open question. At least, the absence of $H$ re-entrant AF above $H_{c2} (0)$ for any field direction implies that the magnetic critical field at $P=0$ will be lower than $H_{c2} (0)$ and already at ambient pressure CeCoIn$_5$ is above $P_c^\star$ \cite{Kne08}.  

\section{The hidden order phase of URu$_2$Si$_2$: feedback between FS instability and spin dynamics}

\begin{figure}[tbh]
\begin{center}
\includegraphics[width=0.7 \linewidth]{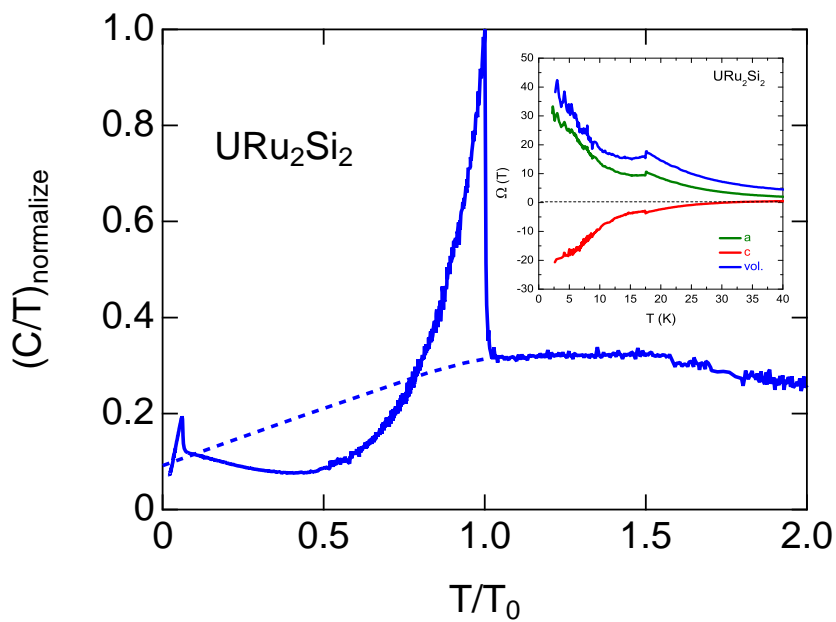}
\end{center}
\caption{Temperature dependence of normalized specific heat for URu$_2$Si$_2$ (full line). The dashed line indicates an extrapolation below $T_0$ of $C/T$ assuming the survival of PM state. The insert shows the $T$ dependence of the Gr\"{u}neisen coefficient \protect\cite{Has08_Sku}.}
\label{fig:fig16}
\end{figure}

Attempts to resolve the nature of the hidden order phase of URu$_2$Si$_2$ boosts an intense experimental and theoretical activity over the last decade (see \cite{Ami07}). 
As several invited contributions (J. C. Davis,  P. M. Oppeneer, and H. Harima (see \cite{Sch10,Elg09,Har10}) have treated this problem, 
we will stress only a few points.
One of the paradox of URu$_2$Si$_2$ represented in Fig.~\ref{fig:fig16} is that
above its ordering temperature $T_0 \sim 17.5\,{\rm K}$, 
it looks as a classical intermediate valence compound with e.g. a broad maximum of $C/T$ as observed for CeSn$_3$~\cite{Has08_Sku}. 
Knowing the entropy involved at low temperature, without any phase transition at $T_0$, 
its residual $\gamma$ term would be around $100\,{\rm mJ\,mole^{-1}K^{-2}}$ 
and a smooth continuous decrease of $C/T$ will occur on cooling. 
However the phase transition at $T_0$ is associated with a drastic change leading  to a marked increase of $C/T$ 
on cooling as observed for usual HFC~\cite{Fis90}. 
Low energy must be involved as the temperature variation of the Gr\"{u}neisen coefficient is large on cooling~\cite{Har_pub3}. 

\begin{figure}[tbh]
\begin{center}
\includegraphics[width=0.7 \linewidth]{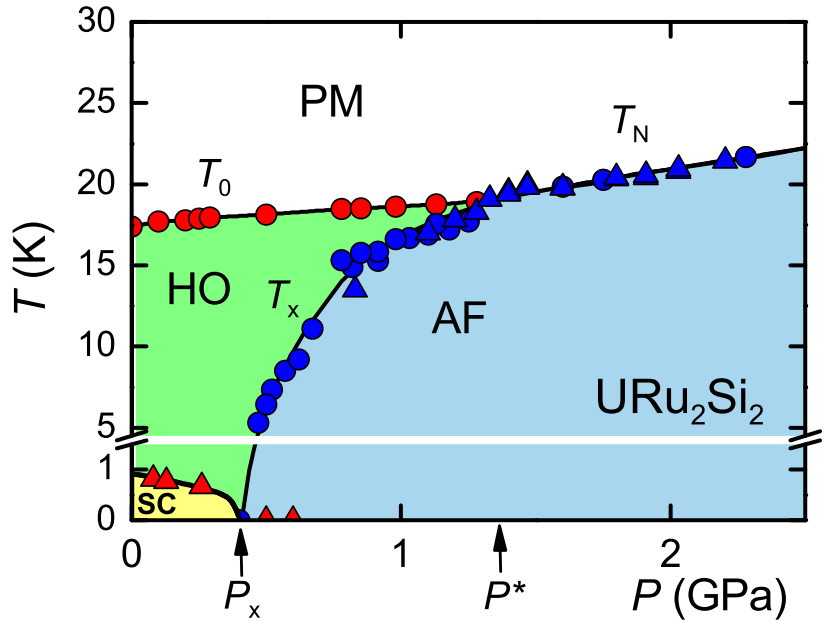}
\end{center}
\caption{($T, P$) phase diagram of URu$_2$Si$_2$ with the location of the PM, HO and HF boundaries. SC exists only in the HO phase i.~e. below $P_x$. AF, HO and PM meets at the initial pressure $P^\ast$ \protect\cite{Has08}. }
\label{fig:fig17}
\end{figure}

It is now well established by transport measurements (Hall effect, Nernst effect, thermoelectric power) (see \cite{Ami07,Has08_Sku}), by NMR~\cite{Mat01},  high energy spectroscopy~\cite{Sant09} and recently by two  scanning tunneling microscope observations~\cite{Sch10,Ayn10} that 
a Fermi surface reconstruction must occur below $T_0$ going from PM to HO states. The ($P, H, T$) phase diagram of URu$_2$Si$_2$ (Fig.~\ref{fig:fig17})~\cite{Has08,Nik10} is now well established despite qualitative differences coming from the high sensitivity of the system to pressure inhomogeneity. 
The observation of an inelastic excitation at wave vector $Q_0 = (1, 0, 0)$ by inelastic neutron scattering experiments in the HO phase at the same wave vector~\cite{Vil08} 
where AF is detected at high pressure via strong elastic neutron reflection and the strong similarity~\cite{Nak03,Has_pub} in the dHvA frequencies detected for both HO and AF ground states are in strong support of the idea that both phases must have the same wave vector $(0, 0, 1)$ implying a doubling of the lattice along $c$. 
This conclusion is reinforced by recent low energy ARPES measurements reported in this conference~\cite{Yos_pub}. 
The idea is that  at $T_0$, a Fermi Surface reconstruction occurs going from body center tetragonal space group {\#}139  above $T_0$ to a simple tetragonal either {\#}136 or {\#}123 (see reference \cite{Har10,Elg09}). 
The persistence inside the tetragonal symmetry is in full agreement with STM data. 
The change of tetragonal space group is associated with a gap opening of characteristic energy $\Delta_{\rm G}$; 
the drop of carrier density has a direct feedback on the spin dynamics 
(development of sharp excitation at $Q_0 = (1, 0, 0)$ and $Q_1 = (0.6, 0, 0)$) which may open the possibility that the U centers even in an intermediate valence phase may have their multipolar properties renormalized to the U$^{4+}$ configuration see \cite{Elg09}).
Thus quadrupole~\cite{Har10} or hexadecapole~\cite{Hau09} are invoked for the OP of URu$_2$Si$_2$. 
It is even proposed that the HO phase may be still a PM state with a lattice switch from space group {\#}139 to {\#}123 \cite{Elg09} with a gap opening in the HO phase  generated by the strong dynamical longitudinal Ising type magnetic fluctuation induced along the $c$ axis. 
An incommensurate spin density wave was proposed in association with the existence of sharp excitations at $Q_1 = (0.6, 0, 0)$~\cite{Wie07}, 
however there is no evidence of any associated detectable OP in neutron scatterings X-ray or STM. 
Despite many attempts the puzzle of the HO is not resolved.
For example, up to now all aims to detect  quadrupole or hexadecapole ordering by X-ray methods at synchrotron beam have failed. 
Recently, careful designed uniaxial strain neutron scattering experiments on URu$_2$Si$_2$~\cite{Bou_pub} indicate that HO and AF cannot be mixed when the strain is applied in the basal plane. 
This result supports strongly the absence of time reversal breaking in HO  opposite to the case of AF. 
Definitively, strong progress has been made but the final solution on the hidden order is not given. 
Thus still different theoretical proposals are made and experimentalists dream to find the solution.

\section{The Future: Material, Instrumentation, Experiments}
\subsection{Material}
The main new trends are often associated with the discovery of unexpected effects on new material. 
In the last decade, we can quote for HFC: 
\begin{description}
\item[-]the ferromagnetic superconductor,
\item[-]the appearance of Ce115 series,
\item[-]the SC in compounds with non-centrosymmetric crystal structure (see E. Bauer this conference ),
\item[-]the criticality of Yb compounds (YbRh$_2$Si$_2$, $\beta$YbAlB$_4$) and recently YbT$_2$Zn$_{20}$ series.
\end{description}
Even on well defined physical problems such as SC of FM HFC it is crucial to find a case at ambient pressure where clean crystals can be grown enabling deep studies on up-up and down-down spin component. 
For Yb HFC, the priority is to find an example with high critical field singularities ($H_{\rm c}$) allowing unambiguously the determination of the Fermi Surface instability.

\subsection{Instrumentation} 
Major progresses have been in experiments in extreme conditions (very low temperature, high pressure, high magnetic fields) which allow large scans for the ($H, P, T$) determination of phase diagram. 
Important breakthroughs have been made in the determination of basic quantities such as specific heat, thermal expansion (via strain gauge, Larmor precession), and NMR. 
High energy spectroscopy can now be realized at low temperature ($T \sim 4\,{\rm K}$) with excellent resolutions. 
STM will certainly be able to give new insights on electronic structures related either with intrinsic properties or extrinsic properties correlated with specific defects. 
One hope is that fancy superstructures or intrinsic defects maybe discovered associated with the strong pressure and uniaxial strain dependence of materials near quantum singularities. 
At least, it is clear that all this improvements already play  an important role in determining the HO state of URu$_2$Si$_2$.

\subsection{Physical problems}
Precise determination of $H, P, T$ phase diagram is a first key goal as studies on Ce-115 HFC or URu$_2$Si$_2$ have shown. 
For example, in ferromagnetic superconductors such as UGe$_2$, 
it is not clear what is the thermodynamic  boundary of SC inside the FM phases and notably, if SC can occur homogeneously inside the low pressure large sublattice magnetization phase FM2. To give a definite answer the combination of experiments  with different probes are absolutely necessary.

An interesting field is the polarized paramagnetic phase which can be obtained through a metamagnetic or pseudo-metamagnetic transition. 
However the full experimental determination of the Fermi surface has never been achieved. 
For example, for CeRu$_2$Si$_2$ above $H_M$, a large part of the Fermi surface is still missing. 
To resolve this enigma is a challenging goal. 
A new emerging phenomena is the occurrence of 2.5 Lifshiftz instability directly associated with the large magnetic polarization induced by the magnetic field (two cases have been pointed out CeIn$_3$~\cite{Gor06} and URu$_2$Si$_2$~\cite{Shi09}.

The detection of other exotic phases such as multipolar ordering~\cite{Kur09}, 
Pomeranchuk instability, nematic phases, Kondo topological insulator~\cite{Dze10} requires the exploration of a large domain.
Systematic studies on skutterudite materials have pointed out quite new cases such as PrFe$_4$P$_{12}$~\cite{Kik07}. 
In unconventional SC, we have already stressed the interest to precisely determine the properties of the AF + SC phases. 
For FM-SC, it is quite obvious that spontaneous vortices must occur as the internal field produced by the magnetization surpasses often the first superconducting critical field $H_{c1} (0)$.
However, up to now a clear observation of  spontaneous vortices have not been achieved~\cite{Oht10}.
Another interesting question concerns (e.g. UCoGe) the possible occurrence of SC on both side of $P_{\rm c}$
may lead to a change in the phases near $P_{\rm c}$~\cite{Min08a}. 
In most HFC-SC, the origin of SC is considered to be mediated through magnetic, 
valence fluctuations or mixed effects of both phenomena. 
In  exotic cases like URu$_2$Si$_2$ or PrOs$_4$Sb$_{12}$~\cite{Bau02}, 
SC is associated  to the HO phase and  to the proximity to a multipolar instability, respectively. 
To clarify their SC singularities with respect to the previous examples is also a stimulating goal.

To stress that many problems remain to be revisited (see \cite{Flo05}), let us indicate that  for example, 
the classification of CeAl$_3$ as a canonical example of PM Kondo lattice is quite controversial as 
(i) in single crystals a clear AF anomaly has been observed \cite{Jac88,Lap93} and 
(ii) microscopic measurements (as muon spectroscopy presented here~\cite{Gav_pub}) confirm that a long range magnetic component exists at $P = 0$. 
These points agree with a negative sign of the thermal expansion indicating that quantum singularity must occur at a slightly higher pressure ($P_{\rm c}\sim 0.2\,{\rm GPa}$, $H_{\rm c}\sim 2\,{\rm T}$) than at ambient pressure. 
For the archetype strong coupling HFC UBe$_{13}$, there is strong evidence that $P$ and $H$ are carrier ``pumps''~\cite{Flo05} 
and thus lead to a continuous change of the zero field SC reference. 
One can speculate that a magnetic field  breaks the cubic symmetry  leading to a drastic change of the FS with $H$. 
This hypothesis have to be tested by modeling  band structure calculations with magnetostriction data. 
Even for the double transition of UPt$_3$, the origin of the postulated symmetry breaking field by AF remains unclear 
as the origin of the tiny ordered moment observed only in neutron scattering experiments 
is far being fully understood. 

Thus, depending on the level of understanding, the HFC are challenging materials with properties which have strong impact on other strongly correlated electronic materials going from organic conductors, high $T_{\rm c}$ SC, recent pnictides, and 3d transition metals.

\section*{References}

\begin{thebibliography}{100}
\bibitem{Jay65}
Jayaraman A 1965 {\em Phys. Rev.\/} {\bf 137} A179--A182

\bibitem{And75}
Andres K, Graebner J~E and Ott H~R 1975 {\em Phys. Rev. Lett.\/} {\bf 43} 1779

\bibitem{Var06}
Varma C 2006 {\em Physica B\/} {\bf 378-380} 17

\bibitem{Tai88}
Taillefer L and Lonzarich G~G 1988 {\em Phys. Rev. Lett.\/} {\bf 60} 1570

\bibitem{Set07}
Settai R, Takeuchi T and {\=O}nuki Y 2007 {\em J. Phys. Soc. Jpn.\/} {\bf 76}
  051003

\bibitem{Ste79}
Steglich F, Aarts J, Bredl C~D, Lieke W, Meschede D, Franz W and {Sch\"{a}fer}
  H 1979 {\em Phys. Rev. Lett.\/} {\bf 43} 1892

\bibitem{Ott83}
Ott H~R, Rudigier H, Fisk Z and Smith J~L 1983 {\em Phys. Rev. Lett.\/} {\bf
  50} 1595

\bibitem{Ste84_UPt3}
Stewart G~R, Fisk Z, Willis J~O and Smith J~L 1984 {\em Phys. Rev. Lett.\/}
  {\bf 52} 679--682

\bibitem{Sch86}
Schlabitz W {\em et al.} 1986 {\em Z. Phys. B\/} {\bf 62} 171

\bibitem{Pal85}
Palstra T~T~M, Menovsky A~A, Berg J~v~d, Dirkmaat A~J, Kes P~H, Nieuwenhuys G~J
  and Mydosh J~A 1985 {\em Phys. Rev. Lett.\/} {\bf 55} 2727.

\bibitem{Fis89_UPt3} 
Fisher R ~A {\em et al.} 1989 {\em Phys. Rev. Lett.\/} {\bf 62} 1411

\bibitem{Cof85}
Coffey L, Rice T~M and Ueda K 1985 {\em J. Phys. C\/} {\bf 18} L813

\bibitem{Pet86}
Pethick C~J and Pines D 1986 {\em Phys. Rev. Lett.\/} {\bf 57} 118

\bibitem{Rav87}
Ravex A, Flouquet J, Tholence J~L, Jaccard D and Meyer A 1987 {\em J. Magn. Magn. Mater.} {\bf 63-64} 400

\bibitem{Jac87}
Jaccard D {\em et al.} 1987 {\em J. Appl. Phys.} {\bf 26} 517.

\bibitem{Col06}
Coleman P 2006 {\em Physica B\/} {\bf 378-380} 1160

\bibitem{Loh07}
L{\"o}hneysen v~H, Rosch A, Vojta M and W{\"o}lfle P 2007 {\em Rev. Mod.
  Phys.\/} {\bf 79} 1015

\bibitem{Flo05}
Flouquet J 2005 {\em Progress in Low Temperature Physics\/} {\bf 15} 139

\bibitem{Miy07}
Miyake K 2007 {\em J. Phys.: Condens. Matter\/} {\bf 19} 12501

\bibitem{Wat09}
Watanabe S, Tsuruta A, Miyake K and Flouquet J 2009 {\em J. Phys. Soc. Jpn.\/}
  {\bf 78} 104706

\bibitem{Flo09}
Flouquet J and Harima H {\em arXiv:0910.3110\/}

\bibitem{Kne06}
Knebel G, Aoki D, Braithwaite D, Salce B and Flouquet J 2006 {\em Phys. Rev.
  B\/} {\bf 74} 020501

\bibitem{Zev88}
Zevin V, Zwicknagl G and Fulde P 1988 {\em Phys. Rev. Lett.\/} {\bf 60}
  2331

\bibitem{Sar96}
Sarrao J {\em et al} 1996 {\em Phys. Rev. B\/} {\bf 54} 12207

\bibitem{Oka10}
Okawa M, Matsunami M, Ishizaka K, Eguchi R, Taguchi M, Chainani A, Takata Y,
  Yabashi M, Tamasaku K, Nishino Y, Ishikawa T, Kuga K, Horie N, Nakatsuji S
  and Shin S 2010 {\em Phys. Rev. Lett.\/} {\bf 104} 247201

\bibitem{Lap81}
Lapierre F {\em et al.} 1981 {\em Solid State Commun.\/} {\bf 40} 347

\bibitem{All79}
Allen J~W, Batlogg B and Wachter P 1979 {\em Phys. Rev. B\/} {\bf 20}
  4807
  
\bibitem{Oka98} 
Okamura H, Kimura S, Shinozaki H, Nanba T, Iga F, Shimizu N and Takabatake T 
1998 {\em Phys. Rev. B\/} {\bf 58} R7496


\bibitem{Bar04}
Barla A {\em et al.} 2004 {\em Phys. Rev. Lett.} {\bf 92} 066401

\bibitem{Imu09}
Imura K, Matsubayashi K, Suzuki H~S, Kabeya N, Deguchi K and Sato N~K 2009 {\em
  J. Phys. Soc. Jpn.\/} {\bf 78} 104602

\bibitem{Der08}
Derr J {\em et al} 2008 {\em Phys. Rev. B\/} {\bf 77} 193197

\bibitem{Tak90}
Takabatake T, Teshima F, Fujii H, Nishigori S, Suzuki T, Fujita T, Yamaguchi Y,
  Sakurai J and Jaccard D 1990 {\em Phys. Rev. B\/} {\bf 41} 9607

\bibitem{Pas10}
Paschen S, Winkler H, Nezu T, Kriegisch M, Hilscher G, Custers J and Prokofiev
  A 2010 {\em J. Phys.: Conf. Ser.\/} {\bf 200} 012156

\bibitem{Kne08}
Knebel G, Aoki D, Brison J~P and Flouquet J 2008 {\em J. Phys. Soc. Jpn.\/}
  {\bf 77} 114704

\bibitem{Mis09}
Misawa T, Yamaji Y and Imada M 2009 {\em J. Phys. Soc. Jpn.\/} {\bf 78} 084707

\bibitem{Mor95}
Moriya T and Takimoto T 1995 {\em J. Phys. Soc. Jpn.\/} {\bf 64} 960-

\bibitem{Bel05_RMP}
Belitz D, Kirkpatrick T~R and Vojta T 2005 {\em Rev. Mod. Phys.\/} {\bf 77} 579

\bibitem{Geg08}
Gegenwart P, Si Q and Steglich F 2008 {\em Nature Physics\/} {\bf 4} 186

\bibitem{Fri09}
Friedemann S, Westerkamp T, Brando M, Oeschler N, Wirth S, Gegenwart P,
  Krellner C, Geibel C and Steglich F 2009 {\em Nature Physics\/} {\bf 5} 465

\bibitem{Jia07}
Jia S, Bud'ko S~L, Samolyuk G~D and Canfield P~C 2007 {\em Nature Physics\/}
  {\bf 3} 334

\bibitem{Hon10}
Honda F, Yasui S, Yoshiuchi S, Takeuchi T, Settai R and \={O}nuki Y 2010 {\em
  J. Phys. Soc. Jpn.\/} {\bf 79} 083709

\bibitem{Suz10}
Suzuki M~T and Harima H 2010 {\em J. Phys. Soc. Jpn.\/} {\bf 79} 024705

\bibitem{Ros88}
Rossat-Mignot, Regnault L~P, Jacoud J~L, Vettier C, Lejay P, Flouquet J, Walker
  E, Jaccard D and Amato A 1988 {\em J. Magn. Magn. Mater.\/} {\bf 76-77} 376

\bibitem{Lac89}
Lacerda A, de~Visser A, Puech L, Lejay P, Haen P, Flouquet J, Voiron J and
  Okhawa F~J 1989 {\em Phys. Rev. B\/} {\bf 40} 11429

\bibitem{Wei10}
Weickert F, Brando M, Steglich F, Gegenwart P and Garst M 2010 {\em Phys. Rev.
  B\/} {\bf 81} 134438

\bibitem{Yan10}
feng Yang Y, Curro N~J, Fisk Z, Pines D and Thompson J~D {\em
  arXiv:1005.5184\/}

\bibitem{Flo04}
Flouquet J {\em et al} 2004 {\em J. Magn. Magn. Mater.\/} {\bf 272-276} 27

\bibitem{MSat01}
Sato M, Koike Y, Katano S, Metoki N, Kadowaki H and Kawarazaki S 2001 {\em J
  .Phys. Soc. Jpn.\/} {\bf Suppl. A 70} 118

\bibitem{Ray98}
Raymond S {\em et al.} 1998 {\em J. Phys.: Condens. Matter\/} {\bf 10} 2363

\bibitem{Mat_pub}
Matsuda T~D {\em et al.} to be published

\bibitem{Flo10}
Flouquet J  to be published in J. Low Temp. Phys.

\bibitem{Shi64}
Shimizu M 1964 {\em Proc. Phys. Soc.\/} {\bf 84} 397

\bibitem{Hux01}
Huxley A, Sheikin I, Ressouche E, Kernavanois N, Braithwaite D, Calemczuk R and
  Flouquet J 2001 {\em Phys. Rev. B\/} {\bf 63} 144519

\bibitem{Tau_pub}
Taufour V {\em et al.} to be published

\bibitem{Kab10}
Kabeya N, Iijima R, Osaki E, Ban S, Imura K, Deguchi K, Aso N, Homma Y,
  Shiokawa Y and Sato N~K 2010 {\em J. Phys.: Conf. Ser.\/} {\bf 200} 032028

\bibitem{Bel99}
Belitz D, Kirkpatrick T~R and Vojta T 1999 {\em Phys. Rev. Lett.\/} {\bf 82}
  4707--4710

\bibitem{Huy07}
Huy N~T, Gasparini A, {de Nijs} D~E, Huang Y, Klaasse J~C~P, Gortenmulder T,
  {de Visser} A, Hamann A, {G\"{o}rlach} T and v~{L\"{o}hneysen} H 2007 {\em
  Phys. Rev. Lett.\/} {\bf 99} 067006

\bibitem{Sax00}
Saxena S~S, Agarwal P, Ahilan K, Grosche F~M, Haselwimmer R~K~W, Steiner M~J,
  Pugh E, Walker I~R, Julian S~R, Monthoux P, Lonzarich G~G, Huxley A, Sheikin
  I, Braithwaite D and Flouquet J 2000 {\em Nature\/} {\bf 406} 587

\bibitem{Aok01}
Aoki D, Huxley A, Ressouche E, Braithwaite D, Flouquet J, Brison J~P, Lhotel E
  and Paulsen C 2001 {\em Nature\/} {\bf 413} 613

\bibitem{Has08_UCoGe}
Hassinger E, Aoki D, Knebel G and Flouquet J 2008 {\em J. Phys. Soc. Jpn.\/}
  {\bf 77} 073703

\bibitem{Has10}
Hassinger E, Aoki D, Knebel G and Flouquet J 2010 {\em J. Phys.: Conf. Ser.\/}
  {\bf 200} 012055

\bibitem{Slo09}
Slooten E, Naka T, Gasparini A, Huang Y~K and de~Visser A 2009 {\em Phys. Rev.
  Lett.\/} {\bf 103} 097003

\bibitem{Har05_pressure}
Hardy F, Huxley A, Flouquet J, Salce B, Knebel G, Braithwaite D, Aoki D, Uhlarz
  M and Pfleiderer C 2005 {\em Physica B\/} {\bf 359} 1111

\bibitem{Miy09}
Miyake A, Aoki D and Flouquet J 2009 {\em J. Phys. Soc. Jpn.\/} {\bf 78} 063703

\bibitem{Har_pub}
Hardy F {\em et al.} to be published

\bibitem{Aok_pub}
Aoki D {\em et al.} to be published

\bibitem{Har09_UGe2}
Hardy F, Meingast C, Taufour V, Flouquet J, v~L{\"o}hneysen H, Fisher R~A,
  Phillips N~E, Huxley A and Lashley J~C 2009 {\em Phys. Rev. B\/} {\bf 80}
  174521

\bibitem{Lev05}
L\'{e}vy F, Sheikin I, Grenier B and Huxley A~D 2005 {\em Science\/} {\bf 309}
  1343

\bibitem{Miy08}
Miyake A, Aoki D and Flouquet J 2008 {\em J. Phys. Soc. Jpn.\/} {\bf 77} 094709

\bibitem{Har_pub2}
Hardy F {\em et al.} to be published

\bibitem{Aok2009}
Aoki D, Matsuda T D, Taufour V, Hassinger E, Knebel G and Flouquet J 2009 {\em J. Phys. Soc. Jpn.\/} {\bf 78} 113709

\bibitem{Min10}
Mineev V P 2010 {\em Phys. Rev. B\/} {\bf 81} 180504


\bibitem{Tad_pub}
Tada Y and Fujimoto S  to be published


\bibitem{Kne04}
Knebel G {\em et al.} 2004 {\em J. Phys.: Condens. Matter\/} {\bf 16} 8905

\bibitem{Par06}
Park T, Ronning F, Yuan H~Q, Salamon M~B, Movshovich R, Sarrao J~L and Thompson
  J~D 2006 {\em Nature\/} {\bf 440} 65
  
\bibitem{Yashima07}
Yashima M, Kawasaki S, Mukuda H, Kitaoka Y, Shishido H, Settai R and \={O}nuki Y 2007 {\em Phys. Rev. B} {\bf 76} 020509


\bibitem{Kne09}
Knebel G, Aoki D and Flouquet J {\em arXiv:0911.5223\/}

\bibitem{Sac09}
Sachdev S {\em arXiv:0910.0846\/}

\bibitem{Ish09}
Ishida K, Nakai Y and Hosono H 2009 {\em J. Phys. Soc. Jpn.\/} {\bf 78} 062001

\bibitem{Bia02}
Bianchi A, Movshovich R, Oeschler N, Gegenwart P, Steglich F, Thompson J~D,
  Pagliuso P~G and Sarrao J~L 2002 {\em Phys. Rev. Lett.\/} {\bf 89} 137002

\bibitem{Mat07}
Matsuda Y and Shimahara H 2007 {\em J. Phys. Soc. Jpn.\/} {\bf 76} 051005

\bibitem{Ful64}
Fulde P and Ferrell R~A 1964  {\em Phys. Rev.\/} {\bf 135} A550

\bibitem{Lar64}
Larkin A~I and Ovchinnikov Y~N 1965 {\em Sov. Phys. JETP\/} {\bf 20} 762

\bibitem{You07}
Young B~L, Urbano R~R, Curro N~J, Thompson J~D, Sarrao J~L, Vorontsov A~B and
  Graf M~J 2007 {\em Phys. Rev. Lett.\/} {\bf 98} 036402

\bibitem{Kou10}
Koutroulakis G, Stewart M~D, Mitrovi\ifmmode~\acute{c}\else \'{c}\fi{} V~F,
  Horvati\ifmmode~\acute{c}\else \'{c}\fi{} M, Berthier C, Lapertot G and
  Flouquet J 2010 {\em Phys. Rev. Lett.\/} {\bf 104} 087001

\bibitem{Ken08}
Kenzelmann M, Str{\"a}ssle T, Niedermayer C, Sigrist M, Padmanabhan B, Zolliker
  M, Bianchi A~D, Movshovich R, Bauer E~D, Sarrao J~L and Thompson J~D 2008
  {\em Science\/} {\bf 321} 1652

\bibitem{Ken10}
Kenzelmann M, Gerber S, Egetenmeyer N, Gavilano J~L, Str\"{a}ssle Th, Bianchi A~D, Ressouche E, Movshovich R, Bauer E~D, Sarrao J~L and Thompson J~D 2010 {\em Phys. Rev. Lett.\/} {\bf 104} 127001

\bibitem{Ron06}
Ronning F, Capan C, Bauer E~D, Thompson J~D, Sarrao J~L and Movshovich R 2006
  {\em Phys. Rev. B\/} {\bf 73} 064519

\bibitem{Tay02}
Tayama T, Harita A, Sakakibara T, Haga Y, Shishido H, Settai R and \={O}nuki Y
  2002 {\em Phys. Rev. B\/} {\bf 65} 180504

\bibitem{How10}
Howald L, Seyfarth G, Knebel G, Lapertot G, Aoki D and Brison J-P 2010 arxiv 1009.4894

\bibitem{Doi10}
Doiron-Leyraud N, Auban-Senzier P, de Cotret S R, Bourbonnais C, J\'erome D, Bechgaard K and Taillefer L 2010 {\em Phys. Rev. B\/} {\bf 80} 214531  

\bibitem{Fuj08}
Fujimoto Y, Tsuruta A and Miyake K 2008 {\em J. Phys. Cond. Matter\/} {\bf 20} 325226

\bibitem{Ike2010}
Ikeda R, Hatayema Y and Aoyama K 2010 {\em Phys. Rev. B} {\bf 82} 060510

\bibitem{Ami07}
Amitsuka H, Matsuda K, Kawasaki I, Tenya K, Yokoyama M, Sekine C, Tateiwa N,
  Kobayashi T~C, Kawarazaki S and Yoshizawa H 2007 {\em J. Magn. Magn.
  Mater.\/} {\bf 310} 214

\bibitem{Sch10}
Schmidt A~R, Hamidian M~H, Wahl P, Meier F, Balatsky A~V, Garrett J~D, Williams
  T~J, Luke G~M and Davis J~C 2010 {\em Nature\/} {\bf 465} 570
  
\bibitem{Elg09}
Elgazzar S, Rusz J, Amft M, Oppeneer P~M and Mydosh J~A 2009 {\em Nature
  Materials\/} {\bf 8} 337

\bibitem{Har10}
Harima H, Miyake K and Flouquet J 2010 {\em J. Phys. Soc. Jpn.\/} {\bf 79}
  033705

\bibitem{Has08_Sku}
Hassinger E, Derr J, Levallois J, Aoki D, Behnia K, Bourdarot F, Knebel G,
  Proust C and Flouquet J 2008 {\em J. Phys. Soc. Jpn. Suppl. A\/} {\bf 77} 172

\bibitem{Fis90}
Fisher R~A {\em et al.} 1990 {\em Physica B} {\bf 163} 419

\bibitem{Har_pub3}
Hardy F {\em et al.} to be published

\bibitem{Mat01}
Matsuda K, Kohori Y, Kohara T, Kuwahara K and Amitsuka H 2001 {\em Phys. Rev.
  Lett.\/} {\bf 87} 087203

\bibitem{Sant09}
Santander-Syro A~F, Klein M, Boariu F~L, Nuber A, Lejay P and Reinert F 2009
  {\em Nature Physics\/} {\bf 5} 637

\bibitem{Ayn10}
Aynajian P, da~Silva~Neto E~H, Parker C~V, Huang Y, Pasupathy A, Mydosh J and
  Yazdani A 2010 {\em PNAS\/} {\bf 107} 10383

\bibitem{Has08}
Hassinger E, Knebel G, Izawa K, Lejay P, Salce B and Flouquet J 2008 {\em Phys.
  Rev. B\/} {\bf 77} 115117

\bibitem{Nik10}
Niklowitz P~G, Pfleiderer C, Keller T, Vojta M, Huang Y~K and Mydosh J~A 2010
  {\em Phys. Rev. Lett.\/} {\bf 104} 106406

\bibitem{Vil08}
Villaume A, Bourdarot F, Hassinger E, Raymond S, Taufour V, Aoki D and Flouquet
  J 2008 {\em Phys. Rev. B\/} {\bf 78} 012504

\bibitem{Nak03}
Nakashima M, Ohkuni H, Inada Y, Settai R, Haga Y, Yamamoto E and \={O}nuki Y
  2003 {\em J. Phys.: Condens. Matter\/} {\bf 15} S2011

\bibitem{Has_pub}
Hassinger E {\em et al.} to be published

\bibitem{Yos_pub}
Yoshida R {\em et al.} to be published and this conference



\bibitem{Hau09}
Haule K and Kotliar G 2009 {\em Nature Physics\/} {\bf 5} 796

\bibitem{Wie07}
Wiebe C~R, Janik J~A, Macdougall G~J, Luke G~M, Garrett J~D, Zhou H~D, Jo Y~J,
  Balicas L, Qiu Y, Copley J~R~D, Yamani Z and Buyers W~J~L 2007 {\em Nature
  Physics\/} {\bf 3} 96

\bibitem{Bou_pub}
Bourdarot F {\em et al.} to be published

\bibitem{Gor06}
Gor'kov L~P and Grigoriev P~D 2006 {\em Phys. Rev. B\/} {\bf 73} 060401

\bibitem{Shi09}
Shishido H, Hashimoto K, Shibauchi T, Sasaki T, Oizumi H, Kobayashi N, Takamasu
  T, Takehana K, Imanaka Y, Matsuda T~D, Haga Y, Onuki Y and Matsuda Y 2009
  {\em Phys. Rev. Lett.\/} {\bf 102} 156403

\bibitem{Kur09}
Kuramoto Y, Kusunose H and Kiss A 2009 {\em J. Phys. Soc. Jpn.\/} {\bf 78}
  072001

\bibitem{Dze10}
Dzero M, Sun K, Galitski V and Coleman P 2010 {\em Phys. Rev. Lett.\/} {\bf
  104} 106408

\bibitem{Kik07} 
  {\bf 76} 043705

\bibitem{Oht10}
Ohta T, Hattori T, Ishida K, Nakai Y, Osaki E, Deguchi K, Sato N~K and Satoh I
  2010 {\em J. Phys. Soc. Jpn.\/} {\bf 79} 023707

\bibitem{Min08a}
Mineev V~P {\em arXiv:0812.2171\/}

\bibitem{Bau02}
Bauer E~D, Frederick N~A, Ho P~C, Zapf V~S and Maple M~B 2002 {\em Phys. Rev.
  B\/} {\bf 65} 100506

\bibitem{Jac88}
Jaccard D, Cibin R, Bezinge A, Sierro J, Matho K, Flouquet J 1988 {\em J. Magn. Magn. Mater.\/} {\bf 76} 255

\bibitem{Lap93}
Lapertot G, Calemczuk R, Marcenat C, Henry J Y, Boucherle J X, Flouquet J, Hammann J, Cibin R, Cors J, Jaccard D, Sierro J 1993 {\em Physica B\/} {\bf 188} 454

\bibitem{Gav_pub}
Gavilano G {\em et al.} to be published and this conference

\end{thebibliography}

\end{document}